\documentclass[preprint]{aastex}
\usepackage{amsmath}
\usepackage{graphicx}
\usepackage{epstopdf}
\usepackage{hyperref}

\shorttitle{$Omh^2(z_1,z_2)$ and $Om(z_1,z_2)$ diagnostics}
\shortauthors{Zheng et al.}

\begin{document}


\title{What are $Omh^2(z_1,z_2)$ and $Om(z_1,z_2)$ diagnostics telling us in light of $H(z)$ data?}


\author{Xiaogang Zheng}
\affil{Department of Astronomy, Beijing Normal University,
    Beijing 100875, China; \\
     Department of Astrophysics and Cosmology, Institute of Physics, University of Silesia, Uniwersytecka 4, 40-007, Katowice, Poland}
\author{Xuheng Ding}
\affil{Department of Astronomy, Beijing Normal University,
    Beijing 100875, China; \\
     Department of Physics and Astronomy, University of California, Los Angeles, CA, 90095-1547, USA}
\author{Marek Biesiada}
\affil{Department of Astronomy, Beijing Normal University,
    Beijing 100875, China; \\
    Department of Astrophysics and Cosmology, Institute of Physics, University of Silesia, Uniwersytecka 4, 40-007, Katowice, Poland}

\and

\author{Shuo Cao, Zong-Hong Zhu}
\affil{Department of Astronomy, Beijing Normal University,
    Beijing 100875, China}


\begin{abstract}

Two-point diagnostics $Om(z_i,z_j)$ and $Omh^2(z_i,z_j)$ have been introduced as an interesting tool for testing the validity of the $\Lambda$CDM model.
Quite recently, \citet{Sahni2014} combined two independent measurements of $H(z)$ from BAO data with the value of the Hubble constant $H_0$, and used the second of these diagnostics to test the $\Lambda$CDM model. Their result indicated a considerable tension between observations and predictions of the
$\Lambda$CDM model. Since reliable data concerning expansion rates of the Universe at different redshifts $H(z)$ are crucial for the successful application
of this method, we investigate both two-point diagnostics on the most comprehensive set of $N=36$ measurements of $H(z)$ coming from the BAO and differential ages (DA) of passively evolving galaxies. We discuss the uncertainties of two-point diagnostics and find that they are strongly non-Gaussian and follow the patterns deeply rooted in their very construction. Therefore we propose that non-parametric median statistics is the most appropriate way of treating this problem. Our results support the claims that $\Lambda$CDM is in tension with $H(z)$ data according to the two-point diagnostics developed by Shafieloo, Sahni and Starobinsky. However, other alternatives to the $\Lambda$CDM, such as wCDM or CPL models perform even worse. We also notice that there are serious systematic differences between BAO and DA methods which ought to be better understood before $H(z)$ measurements can become competitive to the other probes.

\end{abstract}

\keywords{cosmology: observations -- dark energy -- methods: statistical }

\section{Introduction}\label{sec:introduction}
Soon after discovery of accelerating expansion of the Universe \citep{Riess98, Perlmutter99}, the $\Lambda$CDM model has been proposed as the simplest explanation of this phenomenon. Since then it has survived increasingly stringent tests not only related to late accelerating phase of expansion but also as a
framework in which precise CMB data acquired up to the present time could be best understood.
However, many researchers raised serious concerns against claiming that the $\Lambda$CDM was an ultimate solution. Firstly, because of conceptual problems like fine tuning, but
also because of some discrepancies like small-scale anomalies or a recently reported tension between Planck and CFHTLens -- see e.g. \citet{Macaulay, Planck2014, Raveri}.
This motivated many researchers to take a challenge of testing the very foundations of the $\Lambda$CDM. For example, \citet{Zunckel2008} formulated ``a litmus test'' for the $\Lambda$CDM model. Others challenged even
more fundamental aspects like the Copernican principle \citep{Uzan12, Clarkson14}.

However, the most popular probe used to test the $\Lambda$CDM model and to seek evidence of evolving cosmic equation of state is the one initiated by \citet{Sahni2008} after they introduced one point $Om(z)$ diagnostic and generalized it to the two-point case $Om(z_1,z_2) \equiv Om(z_1) - Om(z_2)$.
Later on they developed this further introducing in \citep{Sahni2012} the improved two-point diagnostic $Omh^2(z_1,z_2)$, which they subsequently used in \citet{Sahni2014} to perform this test on three accurately measured values of $H(z)$ from BAO. These were: the $H(z=0)$ measurement by \citep{Riess2011, Planck2014}, $H(z=0.57)$ measurement from SDDS DR9 \citep{Samushia} and the most recent $H(z=2.34)$ measurement from the $Ly{\alpha}$ forest in SDSS DR11 \citep{Delubac}. They found that all three values of the two-point diagnostics $Omh^2(z_1,z_2)$ were in strong tension with the $\Omega_{m,0}h^2$ reported by Planck \citep{Planck2014}.  It has also been noticed \citep{Sahni2014, Delubac} that the $Ly{\alpha}$ forest measurement at $z=2.34$
could be in tension not only with the $\Lambda$CDM model but also with other dark energy models based on the General Relativity. Because such conclusion could be of a paramount importance for dark energy studies, in our recent paper \citep{Ding2015}, we performed this test with a larger sample of $H(z)$ comprising 6 BAO measurements and 23 data-points from cosmic chronometers (differential ages of passively evolving galaxies -- DA hereafter). Essentially, the conclusion was that the tension between $H(z)$ data and $\Lambda$CDM exists. In this paper we study the performance of $Omh^2(z_1,z_2)$ and $Om(z_1,z_2)$ two-point diagnostics in more detail. In section 2 we briefly review the concepts of $Omh^2(z_1,z_2)$ and $Om(z_1,z_2)$. Section 3 reviews the $H(z)$ data. Detailed analysis of statistical properties of both two-point diagnostics and the results obtained with them on $H(z)$ data are subject of the Section 4. Finally we conclude in Section 5.

\section{$Om(z)$ methodology in brief} \label{sec:method}

The so called $Om(z)$ diagnostic has been introduced as an alternative to a common approach of testing models of accelerated expansion of the Universe by phenomenological assumption of a perfect fluid with an equation of state
$p = w \rho $ filling the Universe (in addition to pressureless matter and now dynamically negligible radiation). Cosmological constant $\Lambda$ corresponds formally to $w=-1$. Model independent ``screening test'' of the validity of spatially flat $\Lambda$CDM proposed by \citet{Sahni2008} stems from a simple but smart observation that Friedmann equation in this model: $H(z)^2 = H_0^2 [ \Omega_{m,0} (1+z)^3 + 1 - \Omega_{m,0} ]$ can be rearranged to
\begin{equation} \label{eqOmz}
Om(z) \equiv \frac{{\tilde h}^2(z)-1}{(1+z)^3 - 1} = \Omega_{m,0}
\end{equation}
where ${\tilde h}(z) \equiv H(z) / H_0$. In the literature this dimensionless expansion rate is sometimes denoted as $E(z)$. We retain the notation reminiscent of the Hubble function $H(z)$ and use tilde when it is normalized by the Hubble constant $H_0$ (present expansion rate). We will also use a similar quantity $h(z) \equiv H(z)/ 100\; km\;s^{-1}\;Mpc^{-1}$. Finishing remarks on the nomenclature conventions, let us recall that for historical reasons it is commonly accepted to use the notation $h \equiv H_0 / 100\; km\;s^{-1}\;Mpc^{-1}$ for dimensionless Hubble constant. What is remarkable about Eq.~(\ref{eqOmz}) is the fact that the left hand side is a function of redshift and the right hand side is a number, so the falsifying power of Eq.~(\ref{eqOmz}) is strong. If we knew, from the observations, the expansion rates at different redshifts we would be able to differentiate between $\Lambda$CDM and other dark energy models (including evolving dark energy). Being very attractive from theoretical point of view this test was not easy to preform because there were no accurate direct measurements of $H(z)$ at the time of its formulation, so the researchers willing to use it were forced to reconstruct $H(z)$ from distance measurements of SNIa and this resulted in an increased uncertainty. Currently we are in much better position having at our disposal considerable amount of $H(z)$ measurements obtained from BAO and differential ages techniques, as will be discussed later. Another issue was that $Om(z)$ diagnostic in the $\Lambda$CDM model should not only be constant but exactly equal to the present matter density parameter $\Omega_{m,0}$ which is not easy to measure directly and its value indirectly inferred from CMB or SNIa data was also a subject of debate.

Therefore \citep{Sahni2012} developed this method further by noticing that the two-point diagnostics:
\begin{align} \label{eqOmz1z2}
Om(z_1,z_2) & \equiv Om(z_1) - Om(z_2) = \frac{{\tilde h}^2(z_1)- 1}{(1+z_1)^3 - 1} - \frac{{\tilde h}^2(z_2)- 1}{(1+z_2)^3 - 1}
\end{align}
should always vanish in the $\Lambda$CDM model:
$$Om(z_i,z_j)_{\Lambda CDM} = 0$$ for all $i,j$. If we just knew the expansion rates at different redshifts,
we would be able to tell whether these data are consistent with the $\Lambda$CDM or not without any need of knowing
matter density parameter. As compared to the original $Om(z)$ diagnostic, this two-point diagnostic has another advantage: a sample of $n$ measurements offers us
$\frac{n(n-1)}{2}$ different values of two point diagnostics. As we will see later, this happens at the prize of creating complex statistical properties of two-point diagnostics.
Moreover, vanishing $Om(z_i,z_j)_{\Lambda CDM}$ is again just the litmus test. If
we want to distinguish between different dark energy models, we need to write down corresponding theoretical expression expected for
the right hand side. For the simplest phenomenology of dark energy with constant equation of state parameter $w = const.$, theoretical
expectation for Eq.~(\ref{eqOmz1z2}) should be
\begin{align}\label{Om_wCDM}
 Om(z_i, z_j)_{(wCDM)} =(1-\Omega_{m,0}) \left[ \frac{(1+z_i)^{3(1+w)}-1}{(1+z_i)^3-1}-\frac{(1+z_j)^{3(1+w)}-1}{(1+z_j)^3-1}\right]
\end{align}

Therefore, assuming the redshift ordering $z_j>z_i$, inequality $Om(z_i, z_j)>0$ implies quintessence ($w>-1$) while
$Om(z_i, z_j)<0$ implies phantom scenario ($w<-1$).
Similarly, for the evolving equation of state \citep{Chevalier01,Linder03} modeled by the
Chevalier - Polarski - Linder (CPL) parametrization, the expression should be
\begin{align}\label{Om_CPL}
& Om(z_i, z_j)_{(CPL)}=(1-\Omega_{m,0}) \nonumber \\
& \biggl[\frac{(1+z_i)^{3(1+w_0+w_a)}exp(-3w_az_i/(1+z_i))-1}{(1+z_i)^3-1}-
 \frac{(1+z_j)^{3(1+w_0+w_a)}exp(-3w_az_j/(1+z_j))-1}{(1+z_j)^3-1}\biggr]
\end{align}

In their quite recent paper, \citet{Sahni2014} used a slightly different version of a two-point diagnostic
\begin{equation} \label{eqOmh2}
Omh^2(z_i,z_j) = \frac{h^2(z_i)- h^2(z_j)}{(1+z_i)^3 - (1+z_j)^3}
\end{equation}
which again should be equal to $\Omega_{m,0}h^2$ in the framework of the $\Lambda$CDM model.
For dark energy with constant equation of state $w = const.$, the theoretical expression of Eq.~(\ref{eqOmh2}) should be
\begin{align} \label{Omh2_wCDM}
 Omh^2(z_i,z_j)_{(wCDM)} = \Omega_{m,0}h^2+(1-\Omega_{m,0})h^2
 \left[\frac{(1+z_i)^{3(1+w)} - (1+z_j)^{3(1+w)}}{(1+z_i)^3-(1+z_j)^3}\right]
\end{align}
 and for the CPL parametrization,
one can expect that

\begin{align} \label{Omh2_CPL}
& Omh^2(z_i,z_j)_{(CPL)} = \Omega_{m,0}h^2 + (1-\Omega_{m,0})h^2 \nonumber \\
& \biggl[((1+z_i)^{3(1+w_0+w_a)}e^{\frac{- 3 w_a z_i}{1+z_i}}-
 (1+z_j)^{3(1+w_0+w_a)}e^{\frac{-3w_a z_j}{1+z_j}})/  
 ((1+z_i)^3 - (1+z_j)^3)\biggr]
\end{align}

\section{Data}\label{sec:data}

Our data comprise 36 measurements of $H(z)$ acquired by means of two different techniques. First part of the data comes from
cosmic chronometers \citep{JimenezLoeb}, i.e. massive, early-type galaxies evolving passively on a timescale longer than their age difference.
Certain features of their spectra, such as $D4000$ break at $4000 \;{\AA}$ indicative of the evolution of their stellar populations enable us to
measure age difference of such galaxies. Hence, we use an abbreviation DA for ``differential ages'' to denote cosmic chronometers technique in short.
The most recent results obtained with this technique on a very rich data from the Baryon Oscillation Spectroscopic Survey (BOSS) Data Release 9 have been published by \citet{Moresco16}. Therefore we use 30 measurements of $H(z)$ via DA technique: 23 are the same we have already used in \citet{Ding2015}, supplemented with  two high redshift DA data-points from \citet{Moresco15} and five more $H(z)$ data from \citet{Moresco16}. Second part of our data comes from the analysis of baryon acoustic oscillations (BAO). The BAO data comprise 6 measurements. Table \ref{tablehz} summarizes our data and provides also reference to original sources.

\begin{table*}[htp]
\begin{center}
{{\scriptsize
 \begin{tabular}{l c c c c c c c} \hline\hline
  z & H(z) & $\sigma_H$ & Method & Reference \\ \cline{1-5}
 0.07 & 69 & 19.6 & DA & \citet{Zhang14}\\
 0.09 & 69 & 12 & DA & \citet{Jimenez03}\\
 0.12 & 68.6 & 26.2 & DA & \citet{Zhang14}\\
 0.17 & 83 & 8 & DA & \citet{Simon05}\\
 0.1791 & 75 & 4 & DA & \citet{Moresco12}\\
 0.1993 & 75 & 5 & DA & \citet{Moresco12}\\
 0.2 & 72.9 & 29.6 & DA & \cite{Zhang14}\\
 0.27 & 77 & 14 & DA & \citet{Simon05}\\
 0.28 & 88.8 & 36.6 & DA & \cite{Zhang14}\\
 0.35 & 82.7 & 8.4 & BAO & \citet{Chuang13} \\
 0.3519 & 83 & 14 & DA & \citet{Moresco12}\\
 0.3802 & 83 & 13.5 & DA & \citet{Moresco16}\\
 0.4 & 95 & 17 & DA & \citet{Simon05}\\
 0.4004 & 77 & 10.2 & DA & \citet{Moresco16} \\
 0.4247 & 87.1 & 11.2 & DA & \citet{Moresco16} \\
 0.44 & 82.6 & 7.8 & BAO & \citet{Blake12} \\
 0.4497 & 92.8 & 12.9 & DA & \citet{Moresco16} \\
 0.4783 & 80.9 & 9 & DA & \citet{Moresco16} \\
 0.48 & 97 & 62 & DA & \citet{Stern10}\\
 0.57 & 92.9 & 7.8 & BAO & \citet{Anderson13} \\
 0.5929 & 104 & 13 & DA & \citet{Moresco12}\\
 0.6 & 87.9 & 6.1 & BAO & \citet{Blake12} \\
 0.6797 & 92 & 8 & DA & \citet{Moresco12}\\
 0.73 & 97.3 & 7 & BAO & \citet{Blake12} \\
 0.7812 & 105 & 12 & DA & \citet{Moresco12}\\
 0.8754 & 125 & 17 & DA & \citet{Moresco12}\\
 0.88 & 90 & 40 & DA & \citet{Stern10}\\
 0.9 & 117 & 23 & DA & \citet{Simon05}\\
 1.037 & 154 & 20 & DA & \citet{Moresco12}\\
 1.3 & 168 & 17 & DA & \citet{Simon05}\\
 1.363 & 160 & 33.6 & DA & \citet{Moresco15} \\
 1.43 & 177 & 18 & DA & \citet{Simon05}\\
 1.53 & 140 & 14 & DA & \citet{Simon05}\\
 1.75 & 202 & 40 & DA & \citet{Simon05}\\
 1.965 & 186.5 & 50.4 & DA & \citet{Moresco15} \\
 2.34 & 222 & 7 & BAO & \citet{Delubac} \\
  \hline \hline
\end{tabular}}
\caption{Data of the Hubble parameter $H(z)$ at different redshifts $z$. $H(z)$
and $\sigma_{H}$ are in units of [km s$^{-1}$ Mpc$^{-1}$].\label{tablehz}}}
\end{center}
\end{table*}

Previous papers  by \citet{Sahni2014} and \citet{Ding2015} suggested a tension between
$Omh^2$ calculated from $H(z)$ data and $\Omega_{m,0}h^2=0.1426\pm0.0025$ from Planck satellite \citep{Planck2014}.
In the first step we will readdress this issue using bigger data set of Table \ref{tablehz}. We will go a step further
considering also $Om(z_i, z_j)$ diagnostic and for this purpose we need to assume the specific value of the Hubble
constant $H_0$. We take the value $H_{0(Planck)}=67.4\pm1.4$ suggested by \citep{Planck2014}.
Moreover, we will also consider two more parametrizations for the dark energy, other than $\Lambda$CDM, namely $wCDM$ and $CPL$.
Therefore in order to calculate theoretically expected values of the $Omh^2(z_i, z_j)$ and $Om(z_i, z_j)$ two point diagnostics
we will use $\Omega_{m,0}$, $H_0$ and equation of state parameters in $wCDM$ and $CPL$ models as reported by \citet{Betoule14} (their Tables 14 and 15).
These parameters have been constrained by a combination of the Planck and WMAP satellite measurements of the
CMB temperature fluctuations used jointly with the characteristic scale of the BAO and
the SN Ia Joint Light Analysis (JLA) compilation.
They are summarized in Table~\ref{tableJLAp}.

\begin{table*}[htp]
\caption{The best-fitted values of parameters for three dark energy models obtained from joint analysis of Planck+WP+BAO+JLA data\citep{Betoule14}.}
\begin{center}
{{\scriptsize
 \begin{tabular}{l c c c c c c} \hline\hline
            & $\Omega_{m,0}$ & $H_0$ & $w$ & $w_0$ & $w_a$  \\ \cline{2-6}
 $\Lambda CDM$ & $0.305\pm0.010$ & $68.34\pm1.03$ & $\square$ & $\square$  & $\square$   \\
 $wCDM$ & $0.303\pm0.012$ & $68.50\pm1.27$ & $-1.027\pm0.055$ &  $\square$ &  $\square$  \\
 $CPL$ & $0.304\pm0.012$ & $68.59\pm1.27$ & $\square$  & $-0.957\pm0.124$ & $-0.336\pm0.552$   \\
 \hline \hline
\end{tabular}} \label{tableJLAp}}
\end{center}
\end{table*}

\begin{figure}[htbp]
\begin{center}
\centering
\includegraphics[width=0.45\textwidth]{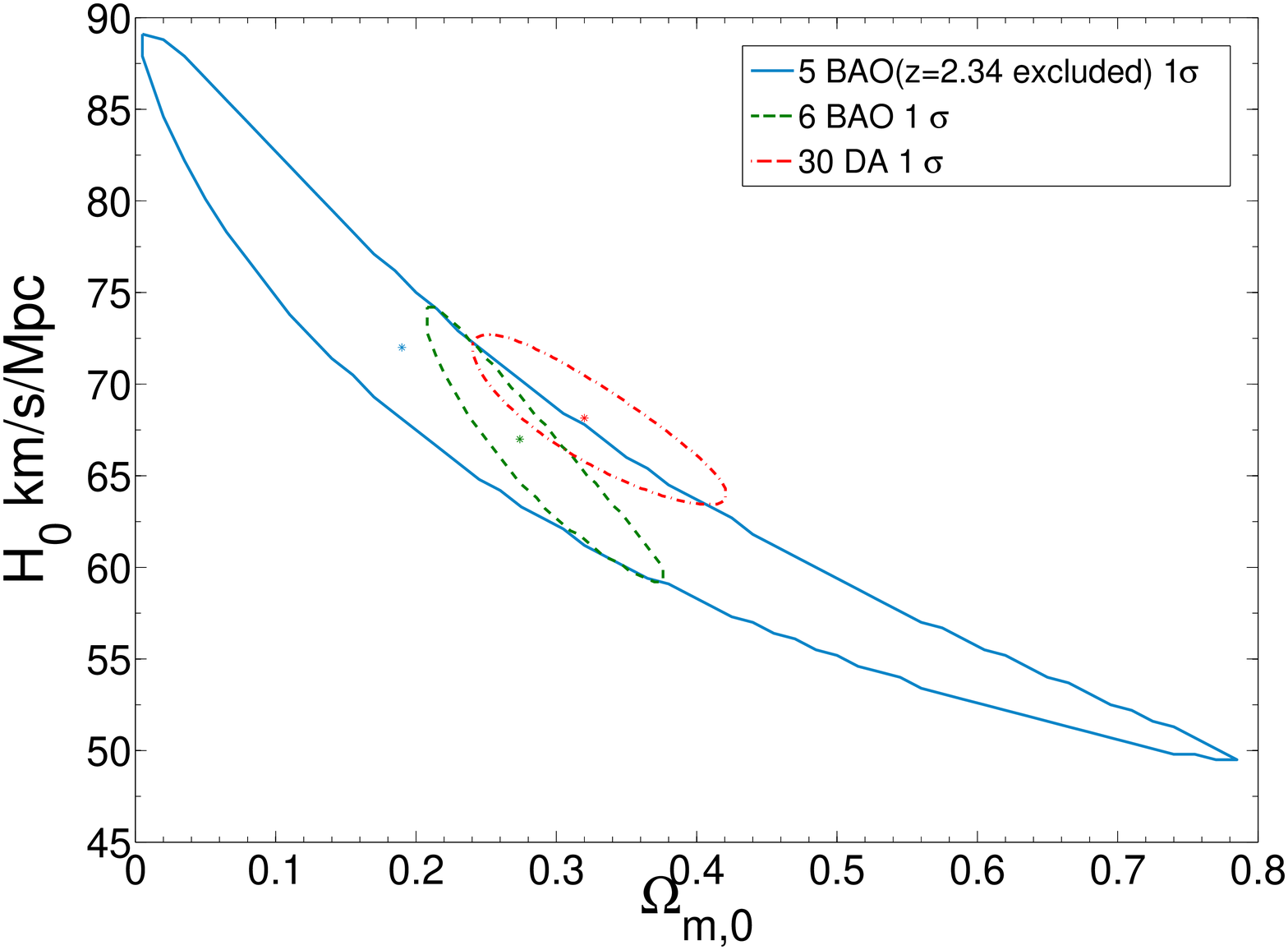}
\caption{Comparison of constraints on the $\Lambda$CDM model parameters from $N=30$ DA data (dash-dot red), $N=6$ BAO data (dash green) and $N=5$ BAO with $z=2.34$ measurement excluded (solid blue). $68\%$ confidence regions are shown with crosses denoting central fits.
}\label{figLCDMconstraint}
\end{center}
\end{figure}

Because the $H(z)$ data set we used is inhomogeneous we performed our analysis of two point diagnostics not only on a full sample of $N=36$ combined DA+BAO measurements, but also on DA ($N=30$) and BAO ($N=6$) separately. Moreover, since the $z=2.34$ measurement \citep{Delubac} turns out to have a big leverage on BAO results we have also considered $N=35$ sub-sample by excluding this measurement from the full DA+BAO sample. The above mentioned leverage can be seen on Fig.~\ref{figLCDMconstraint} where we have used different samples of $H(z_i)$ to constrain the $(\Omega_{m,0}, H_0)$ parameters in the spatially flat $\Lambda$CDM model where $H(z)=H_0\sqrt{\Omega_{m,0}(1+z)^3 + 1-\Omega_{m,0}}$. One can see that inclusion of $z=2.34$ data point improves dramatically
the BAO fit but still there is a mismatch between BAO and DA $68\%$ confidence regions.

\section{Results}\label{sec:results}

\begin{figure*}
\begin{center}
\includegraphics[width=1\textwidth]{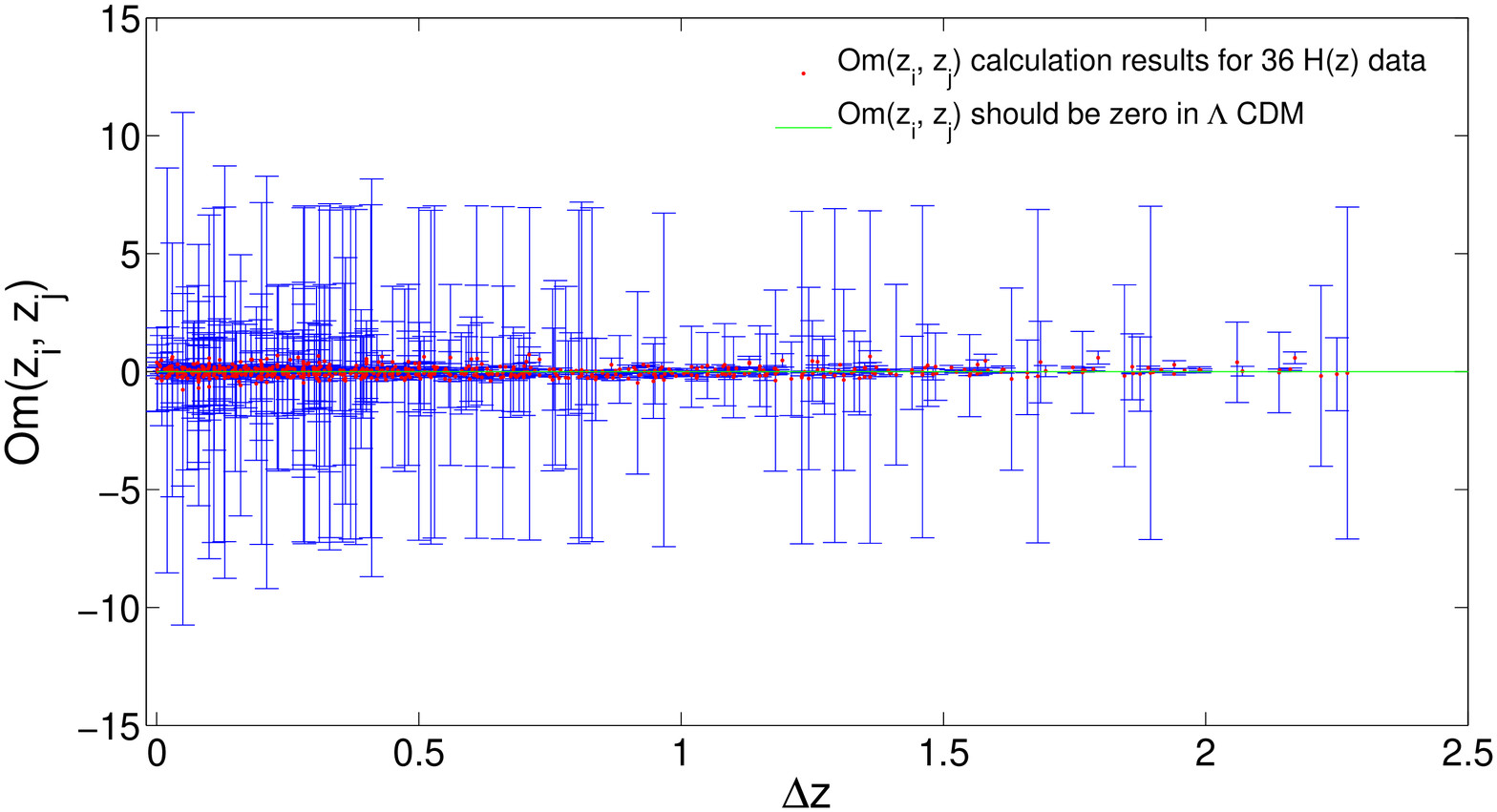}
\includegraphics[width=1\textwidth]{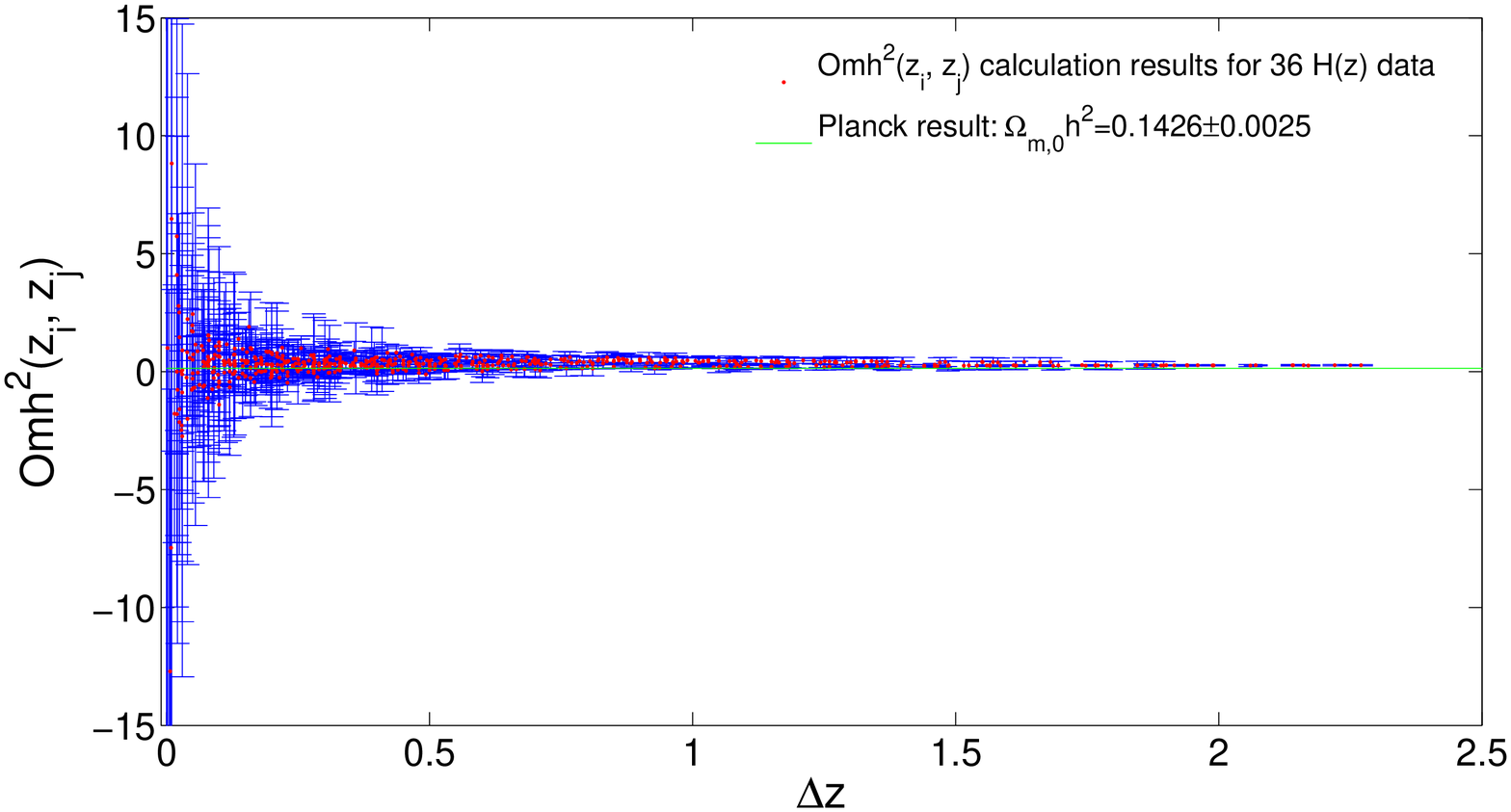}
\caption{The $Om(z_i, z_j)$ ({\it Upper panel}) and $Omh^2(z_i, z_j)$ ({\it Lower panel}) two point diagnostics calculated on the
full sample of 36 $H(z)$ data. Red points denote the calculated central values (weighted means)
and the blue bars ---  corresponding uncertainties, green line denote the values to which two point diagnostic is expected to be equal to within the $\Lambda$CDM
model: zero for the $Om(z_i, z_j)$ and $\Omega_{m,0}h^2=0.1426\pm0.0025$ for the $Omh^2(z_i, z_j)$.
}
\label{figpair}
\end{center}
\end{figure*}

In order to gain insight concerning the $Omh^2(z_i, z_j)$ and $Om(z_i, z_j)$ two point diagnostics calculated for every combination of pairs taken from 36 $H(z)$ data points, i.e. totally 630 pairs of $(z_i, z_j)$, Fig. ~\ref{figpair} displays these diagnostics together with their
uncertainties as a function of redshift difference $\Delta z = |z_i - z_j|$. There are some interesting features regarding the uncertainties of the two point diagnostics. One can see that they are apparently non-gaussian and two points diagnostics -- especially $Omh^2(z_i, z_j)$ -- are heteroscedastic.
Reasons for this can be understood by looking at the formulae for the corresponding uncertainties. Namely, applying the error propagation formula to the definitions of $Omh^2(z_i, z_j)$, i.e. Eq.~(\ref{eqOmh2}), and $Om(z_i, z_j)$, i.e. Eq.~(\ref{eqOmz1z2}), one obtains respectively:
\begin{equation}\label{eqOmh2sig}
\sigma^2_{Omh^2,ij}=\frac{4 \left( h^2(z_i)\sigma^2_{h(z_i)}+h^2(z_j)\sigma^2_{h(z_j)} \right)}{\left( (1+z_i)^3-(1+z_j)^3 \right)^2}
\end{equation}
where $\sigma_{h(z_i)}$ denotes the uncertainty of the $i-th$ Hubble parameter measurement in units of $[100 km\; s^{-1} \; Mpc^{-1}]$, i.e. $\sigma_{h(z_i)} = 0.01 \; \sigma_{H(z_i)}$, and:
\begin{equation}\label{eqomsig}
\sigma^2_{Om,ij}=\frac{4{\tilde h}^2(z_i)\sigma^2_{{\tilde h}(z_i)}}{\left( (1+z_i)^3-1 \right)^2}+\frac{4{\tilde h}^2(z_j)\sigma^2_{{\tilde h}(z_j)}}{\left( (1+z_j)^3-1 \right)^2}
\end{equation}
where in this case, because of normalizing to the actual Hubble constant $H_0$ one has:
\begin{equation}\label{eqomhsig}
\sigma^2_{\tilde h(z_i)}= \left( \frac{\sigma_{H(z_i)}}{H_0} \right)^2+ \left( \frac{H(z_i)\sigma_{H_0}}{H_0^2} \right)^2
\end{equation}
and $\sigma_{H(z_i)}$ denotes the uncertainty of the $i-th$ Hubble parameter measurement
in units of $[km\; s^{-1} \; Mpc^{-1}]$.
Now one can see from Eq.~(\ref{eqOmh2sig}) that the uncertainty of $Omh^2(z_i, z_j)$ is large whenever the redshifts $z_i$ and $z_j$ are close to each other, whereas the uncertainty of $Om(z_i, z_j)$ is large whenever one of the redshifts in the pair is close to zero.

Two point diagnostics used as tests of the $\Lambda$CDM model are supposed to give just a constant numerical value for this model
therefore one should first make a summary statistics of their values calculated on the data sets.
Because of the statistical properties discusses above, we used two approaches.
First was to calculate the weighted mean since it is the most popular way of summarizing measurements encountered in the literature,
unfortunately sometimes without checking the validity of such an approach.
The weighted mean formula for the $Om(z_i, z_j)$ diagnostic reads:
\begin{equation}
Om_{(w.m.)}=\frac{\sum^{n-1}_{i=1}\sum^{n}_{j=i+1}Om(z_i,z_j)/\sigma^2_{Om,ij}}{\sum^{n-1}_{i=1}\sum^{n}_{j=i+1}1/\sigma^2_{Om,ij}}
\end{equation}\label{eqOmwm}
and its variance is:
\begin{equation}
\sigma^2_{Om_{(w.m.)}}= \left( \sum^{n-1}_{i=1}\sum^{n}_{j=i+1}1/\sigma^2_{Om,ij} \right)^{-1}
\end{equation}\label{eqOmwmsig}
with $\sigma^2_{Om,ij}$ given by Eq.~(\ref{eqomsig}).
Similarly, the weighted mean formula for $Omh^2(z_i, z_j)$ diagnostic is:
\begin{equation}
Omh^2_{(w.m.)}=\frac{\sum^{n-1}_{i=1}\sum^{n}_{j=i+1}Omh^2(z_i,z_j)/\sigma^2_{Omh^2,ij}}{\sum^{n-1}_{i=1}\sum^{n}_{j=i+1}1/\sigma^2_{Omh^2,ij}}
\end{equation}\label{eqOmh2wm}
and its variance is:
\begin{equation}
\sigma^2_{Omh^2_{(w.m.)}}= \left( \sum^{n-1}_{i=1}\sum^{n}_{j=i+1}1/\sigma^2_{Omh^2,ij} \right)^{-1}
\end{equation}\label{eqOmh2wmsig}
with $\sigma^2_{Omh^2,ij}$ given by Eq.~(\ref{eqOmh2sig}).

The second approach is the ``Median Statistics'' which was pioneered by \citet{GottIII}.
It is based on a very well known property of the median that being a non-parametric measure is robust
and can be used without any prior assumption about the underlying distribution, in particular without
assuming its Gaussianity. From the definition of the median, probability that any particular measurement, one
of N independent measurements is higher than the true median is $50\%$. Consequently,
the probability that $n$-th observation out of the total number of $N$ is higher than the
median follows the binomial distribution: $P=2^{-N}N!/[n!(N-n)!]$. This property allows to
calculate the $68\%$ confidence intervals (CI) of the median.

\begin{table*}[htp]
\begin{center}
\caption{Results of $Om(z_i,z_j)$ and $Omh^2(z_i,z_j)$ two point diagnostics calculated on different sub-samples using the weighted mean and the median statistics.
For the $Om(z_i,z_j)$ diagnostic the Hubble constant value of $H_0=67.4\pm1.4 \; km s^{-1} Mpc^{-1}$ was assumed.
The results of the $Omh^2(z_i,z_j)$ diagnostic should be compared to
Planck result $\Omega_{m,0}h^2_{(Planck)}=0.1426\pm0.0025$. The percentage of residuals distribution falling within $|N_{\sigma}|<1$ for the main sample and different sub-samples is shown as an indicator of non-Gaussianity.}
\begin{tabular}{l c c c c } \hline\hline
     & $Om(z_i, z_j)_{(w.m.)}$ & $|N_{\sigma}|<1$ & $Om(z_i, z_j)_{(m.s.)}$ & $|N_{\sigma}|<1$ \\ \cline{1-5}
Full sample (n=36) & $-0.0061\pm0.0111$ & $91.90\%$ & $-0.0199^{+0.0077}_{-0.0089}$ & $92.22\%$\\
z=2.34 excluded (n=35) & $-0.0137\pm0.0123$ & $92.61\%$ & $-0.0259^{+0.0090}_{-0.0046}$ & $92.61\%$\\
DA only (n=30) & $-0.0019\pm0.0165$ & $92.87\%$ & $-0.0305^{+0.0077}_{-0.0129}$ & $93.10\%$\\
BAO only (n=6) & $0.0058\pm0.0351$ & $100\%$ & $0.0326^{+0.0093}_{-0.0063}$ & $100\%$ \\
\hline\hline
    & $Omh^2(z_i, z_j)_{(w.m.)}$ & $|N_{\sigma}|<1$ &$Omh^2(z_i, z_j)_{(m.s.)}$ & $|N_{\sigma}|<1$ \\ \cline{1-5}
Full sample (n=36) & $0.1259\pm0.0019$ & $83.49\%$ & $0.1501^{+0.0049}_{-0.0082}$ & $79.37\%$\\
z=2.34 excluded (n=35) & $0.1404\pm0.0040$ & $82.02\%$ & $0.1586^{+0.0029}_{-0.0048}$ & $85.04\%$ \\
DA only (n=30) & $0.1437\pm0.0046$ & $81.61\%$ & $0.1729^{+0.0027}_{-0.0076}$ & $87.82\%$\\
BAO only (n=6) & $0.1231\pm0.0045$ & $100\%$ & $0.1218^{+0.0002}_{-0.0011}$ & $100\%$\\
\hline\hline
\end{tabular}\label{tableomh2}
\end{center}
\end{table*}

\begin{figure*}
\begin{center}
\includegraphics[angle=0,width=40mm]{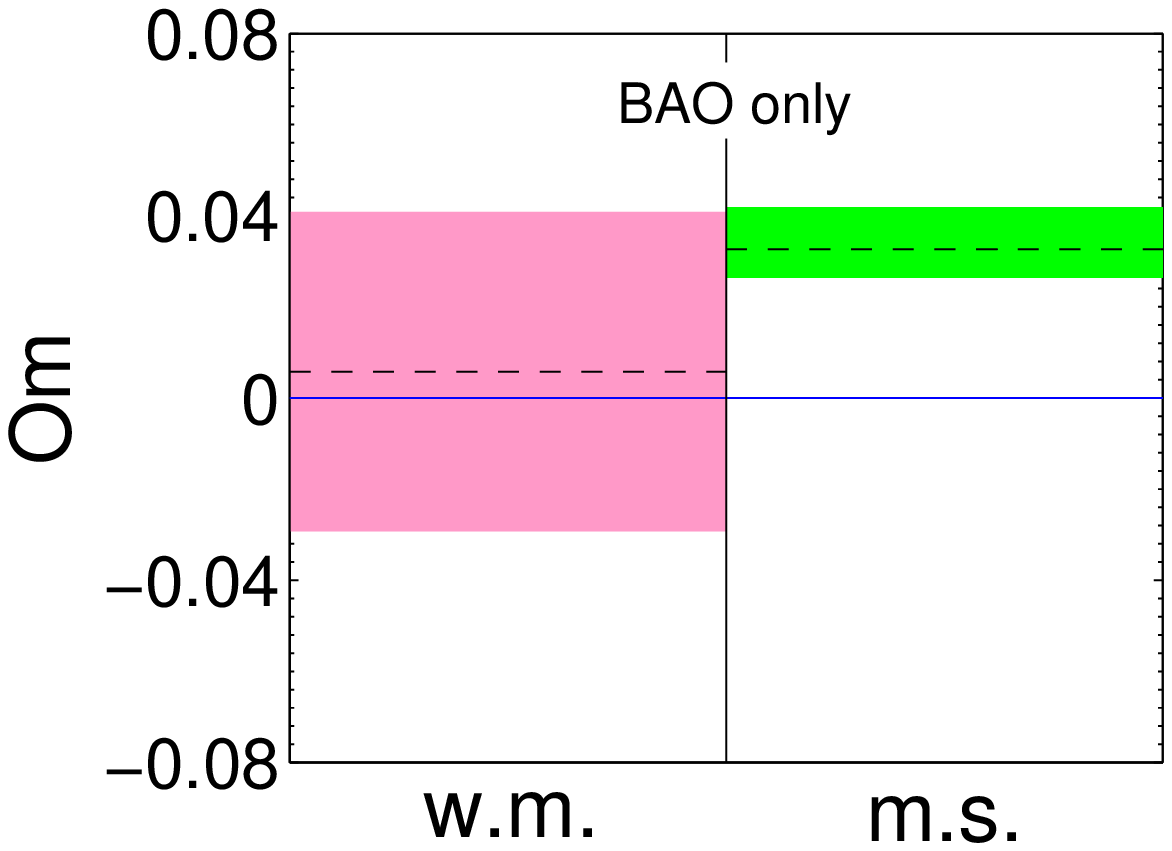}
\includegraphics[angle=0,width=40mm]{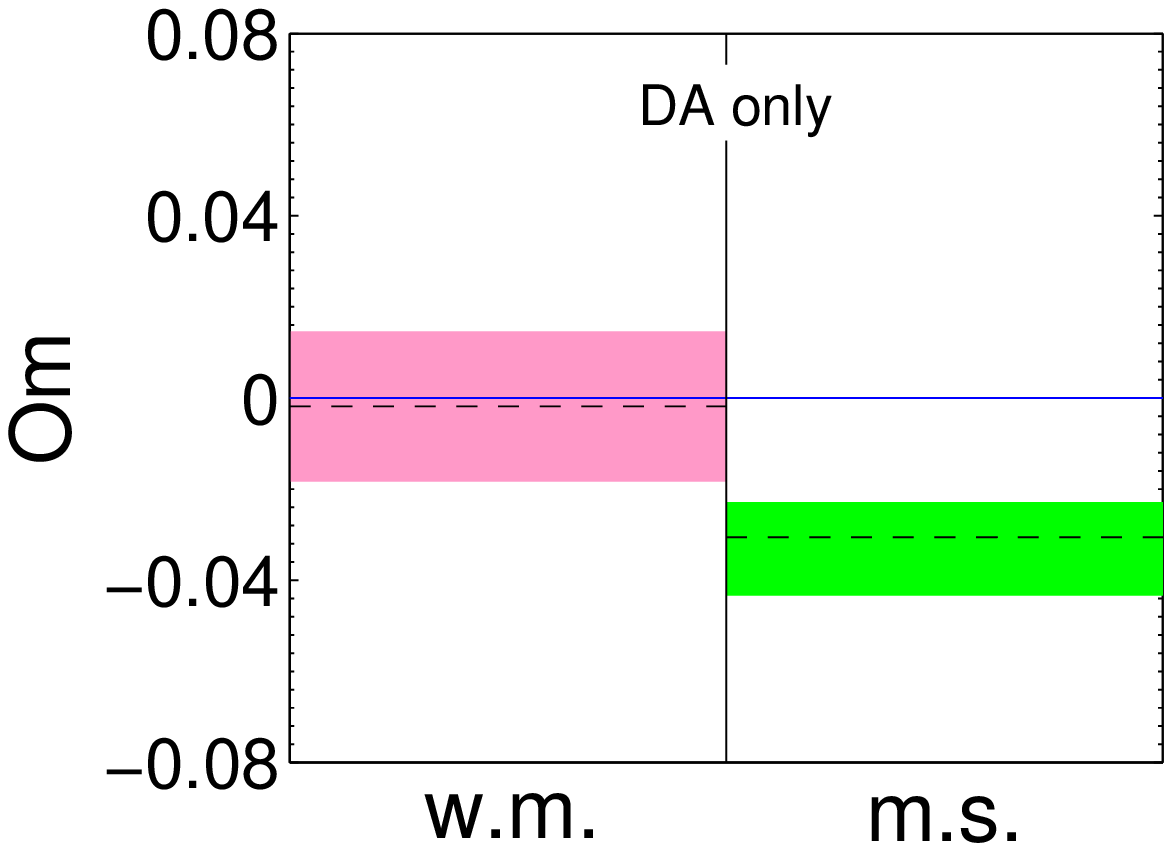}
\includegraphics[angle=0,width=40mm]{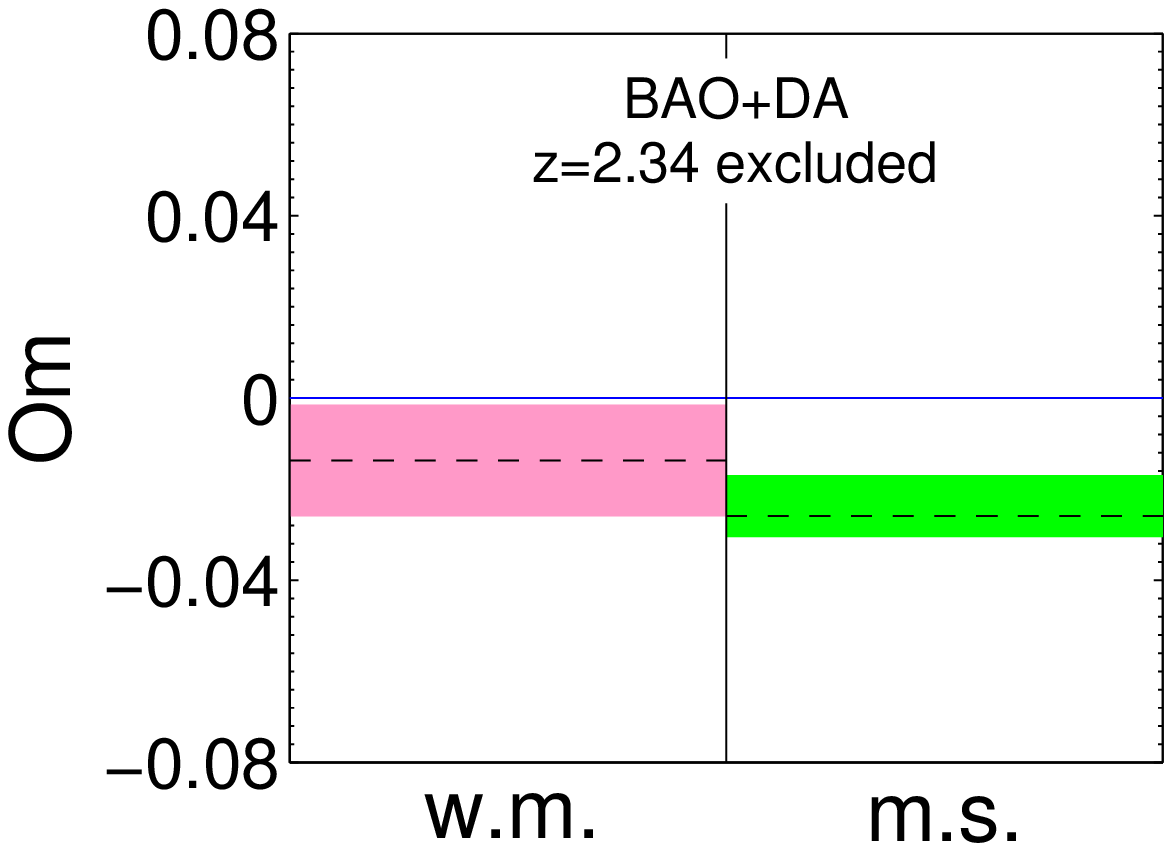}
\includegraphics[angle=0,width=40mm]{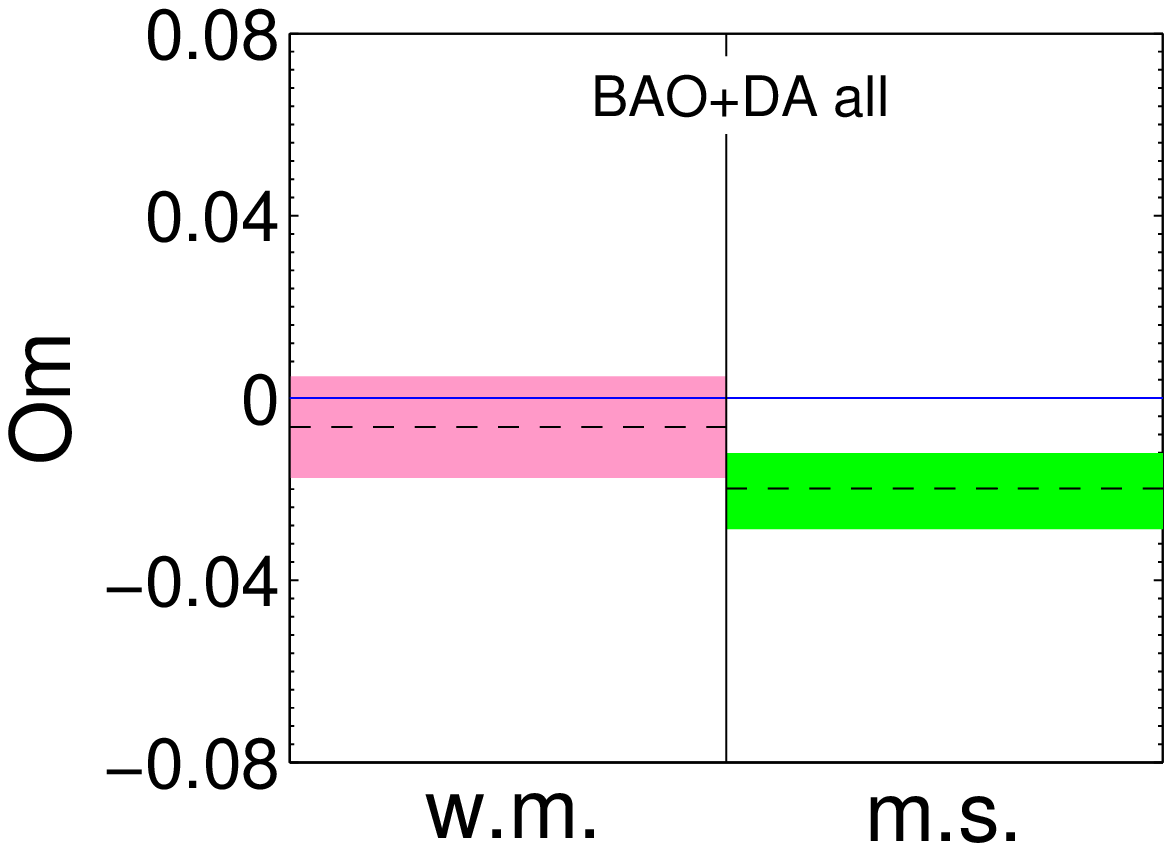}
\caption{The $Om(z_i, z_j)$ two point diagnostic displayed as the weighted mean (left panels) and as the median value (right panels) indicated by dashed lines surrounded by color bands denoting $68\%$ confidence regions.
Long solid line shows the $Om(z_i, z_j)=0$ level expected for the $\Lambda$CDM.  Four figures correspond to four respective sub-samples: $N=6$ BAO data, $N=30$ DA data, $N=35$ combined BAO+DA sample with the $H(z=2.34)$
data point excluded and the full $N=36$ combined BAO+DA data.}
\label{figom}
\end{center}
\end{figure*}

\begin{figure*}
\begin{center}
\includegraphics[angle=0,width=40mm]{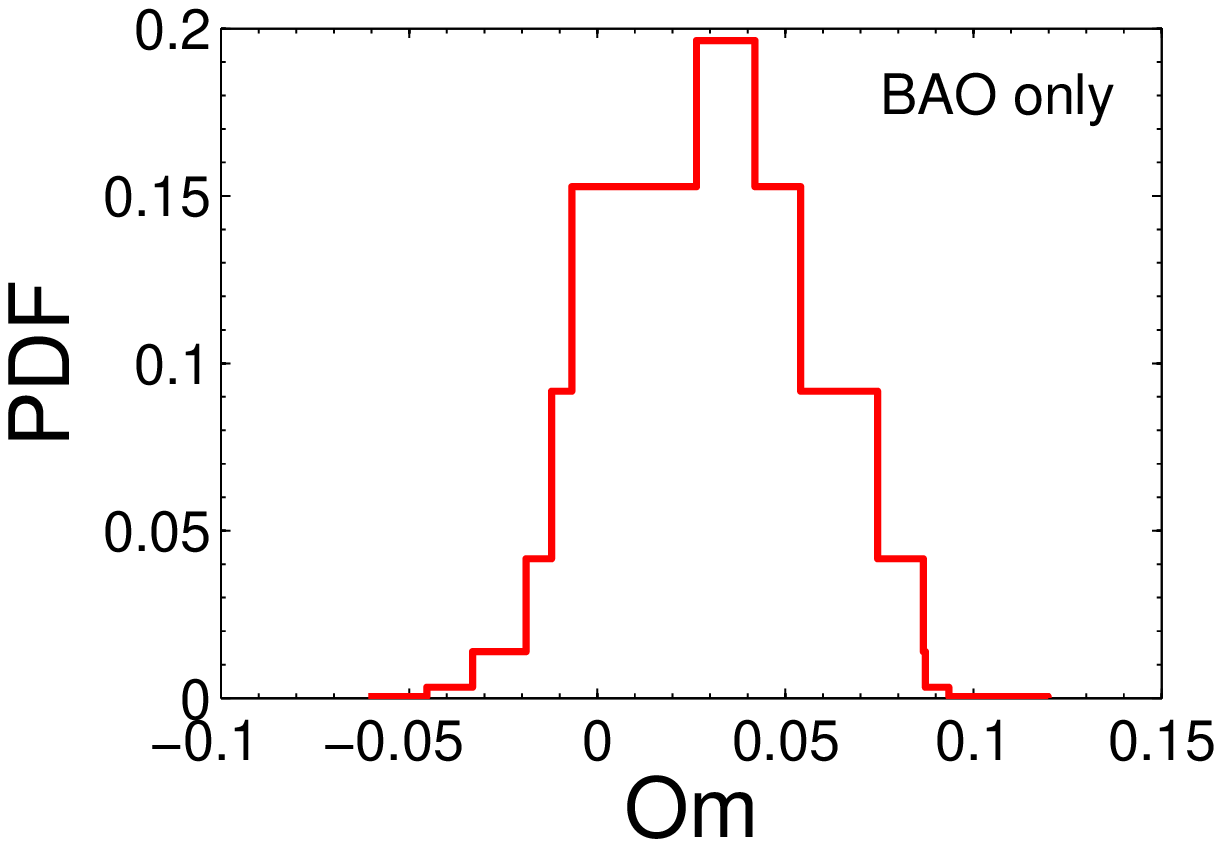}
\includegraphics[angle=0,width=40mm]{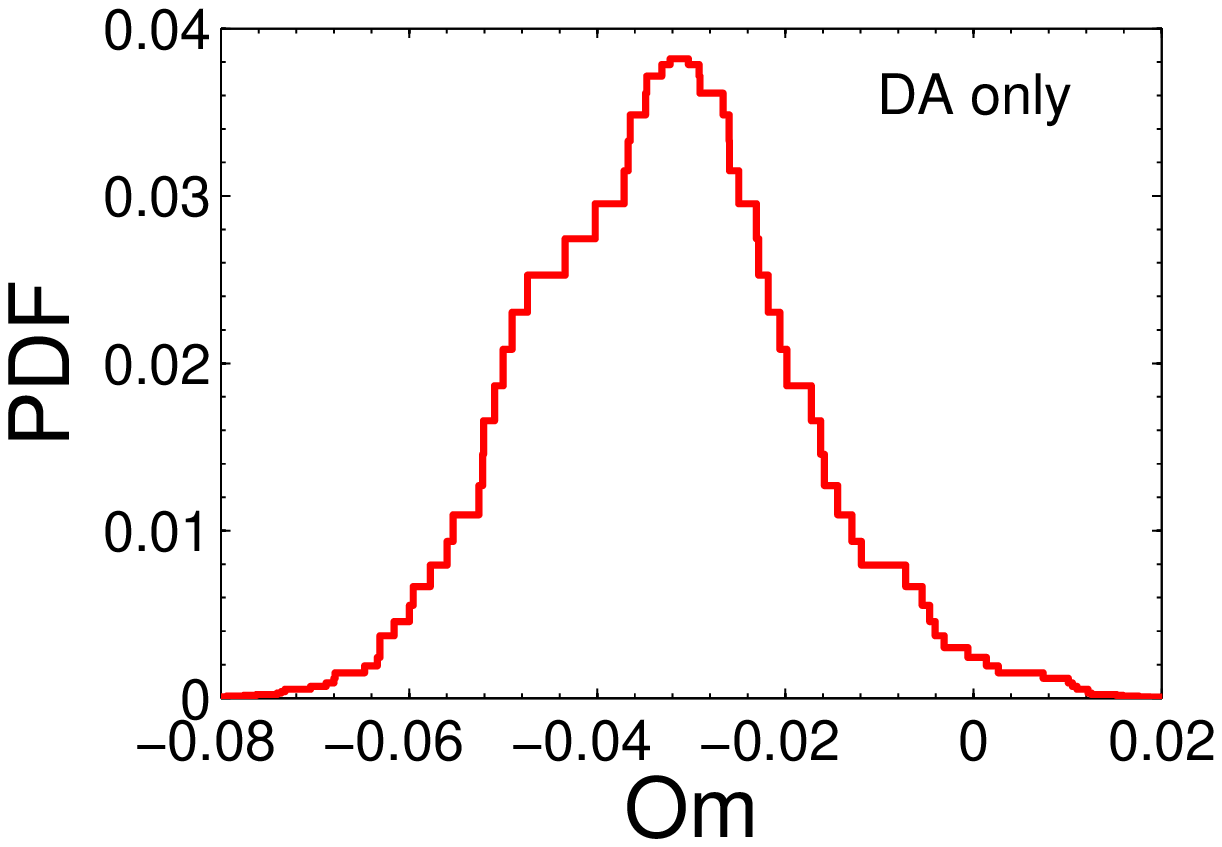}
\includegraphics[angle=0,width=40mm]{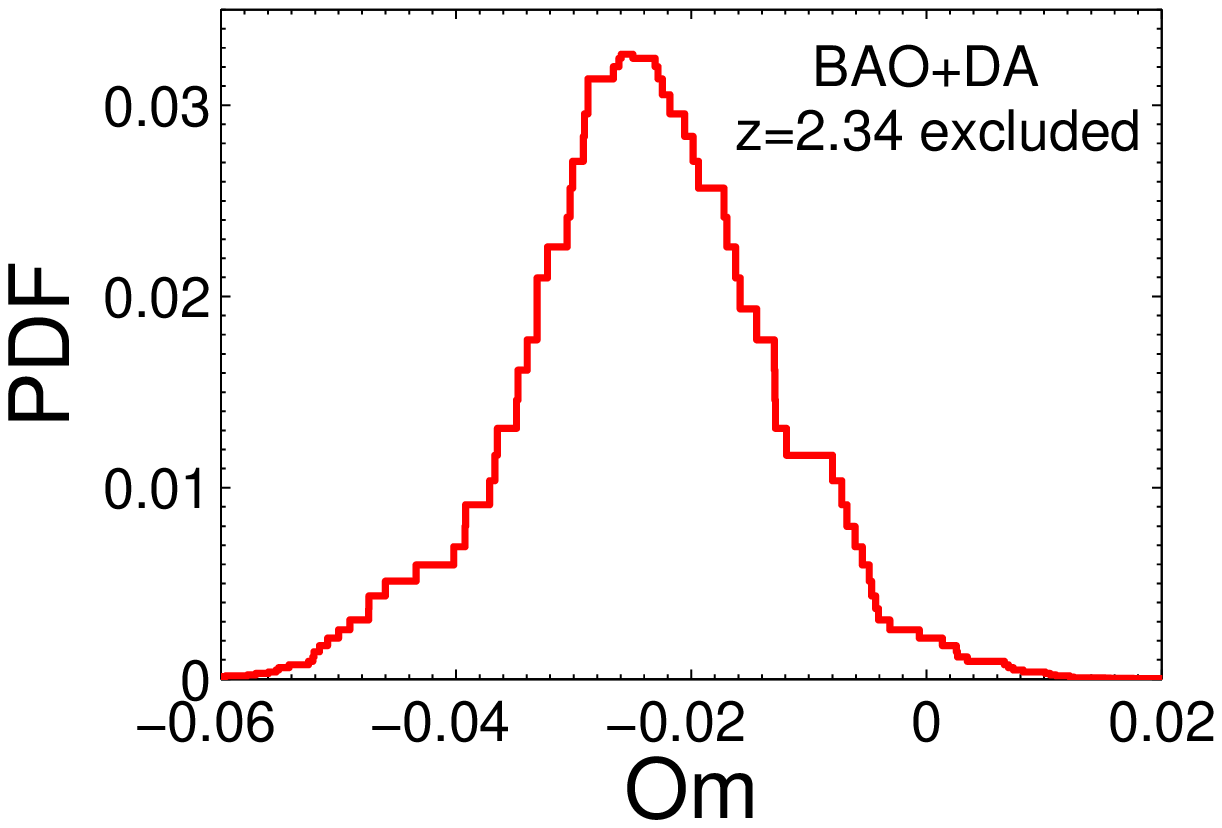}
\includegraphics[angle=0,width=40mm]{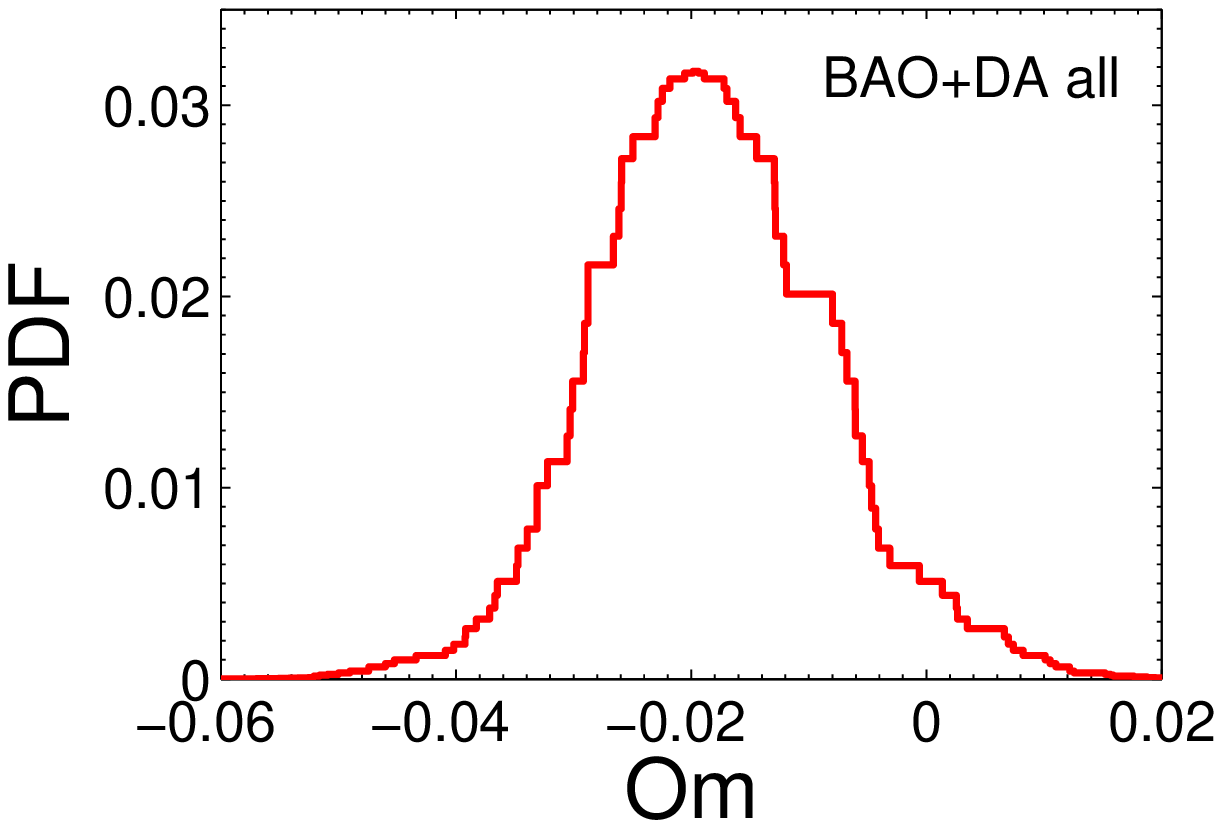}
\caption{Histograms of the $Om(z_i, z_j)$ two point diagnostic calculated with different samples: $N=6$ BAO data, $N=30$ DA data, $N=35$ combined BAO+DA sample with the $H(z=2.34)$
data point excluded and the full $N=36$ combined BAO+DA data.}
\label{figPDFobom}
\end{center}
\end{figure*}

\begin{figure*}
\begin{center}
\includegraphics[angle=0,width=40mm]{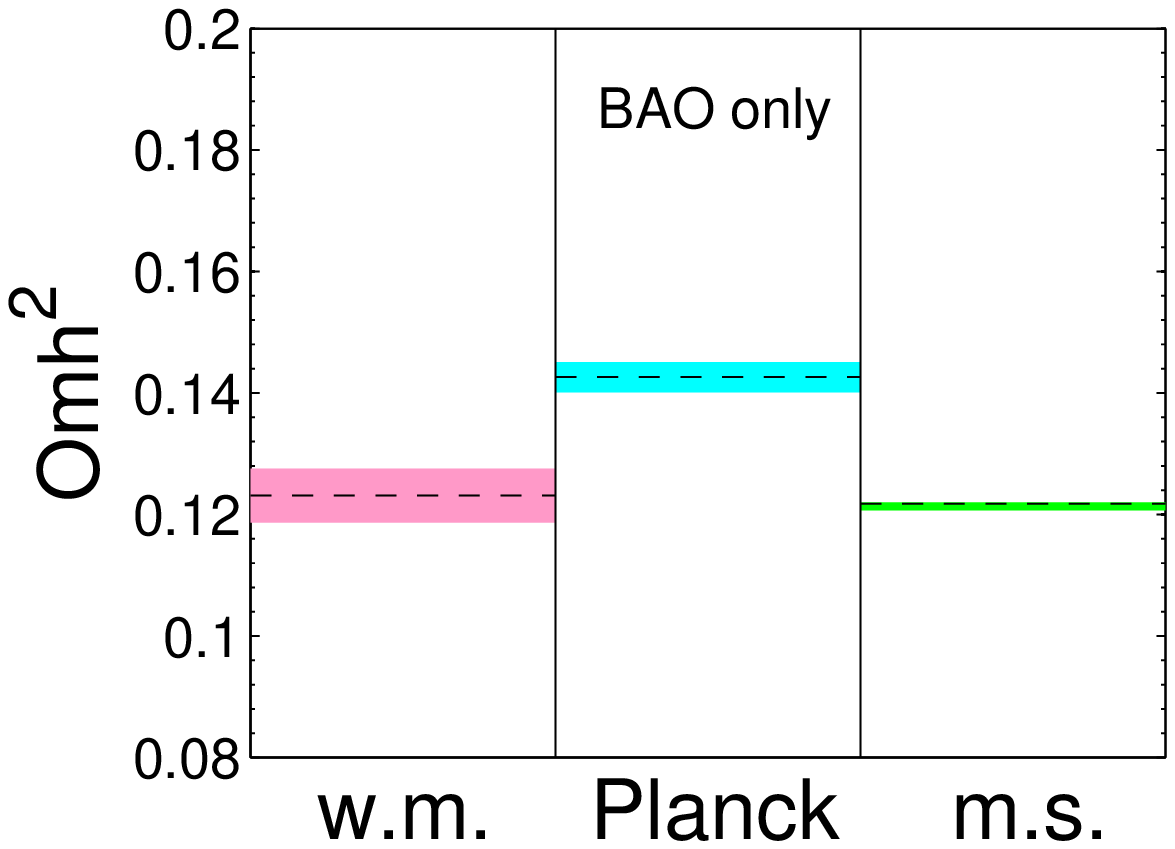}
\includegraphics[angle=0,width=40mm]{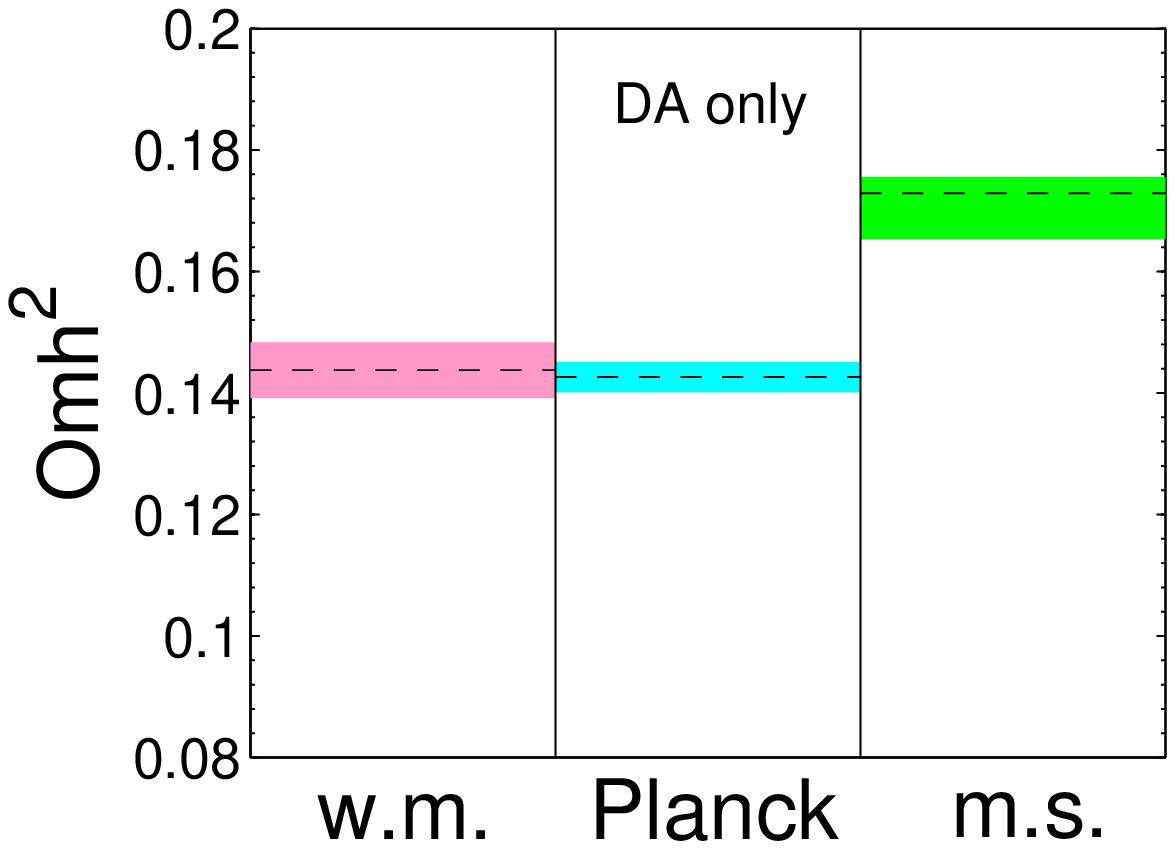}
\includegraphics[angle=0,width=40mm]{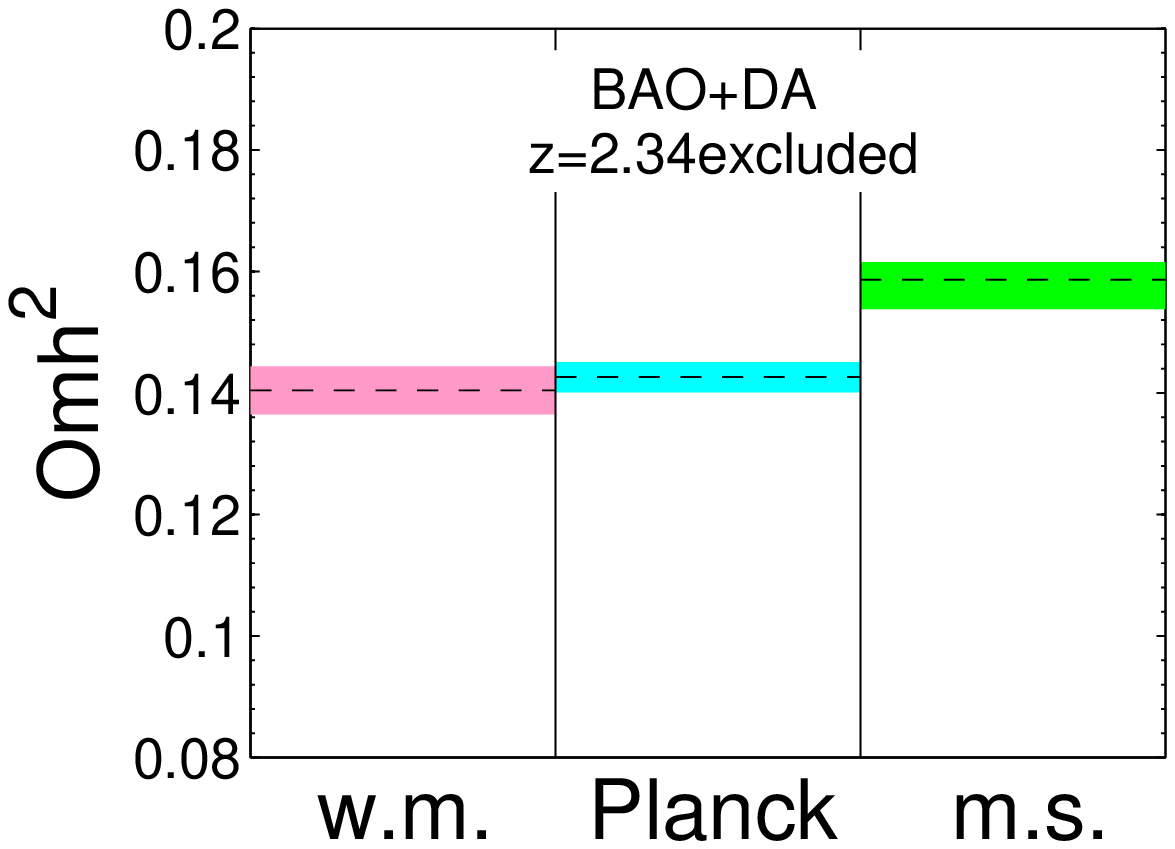}
\includegraphics[angle=0,width=40mm]{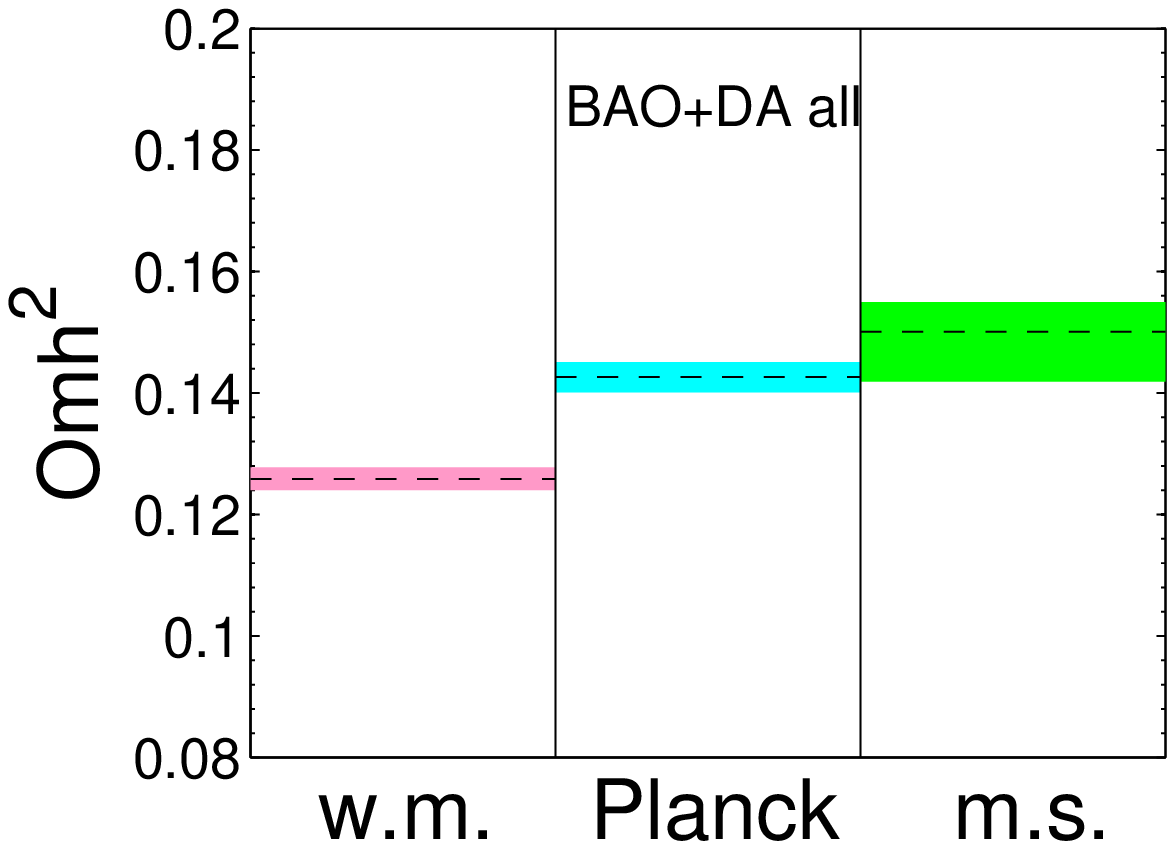}
\caption{The $Omh^2(z_i, z_j)$ two point diagnostic displayed as the weighted mean (left panels) and as the median value (right panels) indicated by dashed lines surrounded by color bands denoting $68\%$ confidence regions. The middle panels show the Planck result of $\Omega_{m,0}h^2_{(Planck)}=0.1426\pm0.0025$ --- this is the value expected for $Omh^2(z_i, z_j)$ within $\Lambda$CDM model.
Four figures correspond to four respective sub-samples: $N=6$ BAO data, $N=30$ DA data, $N=35$ combined BAO+DA sample with the $H(z=2.34)$
data point excluded and the full $N=36$ combined BAO+DA data.}
\label{figomh2}
\end{center}
\end{figure*}

\begin{figure*}
\begin{center}
\includegraphics[angle=0,width=40mm]{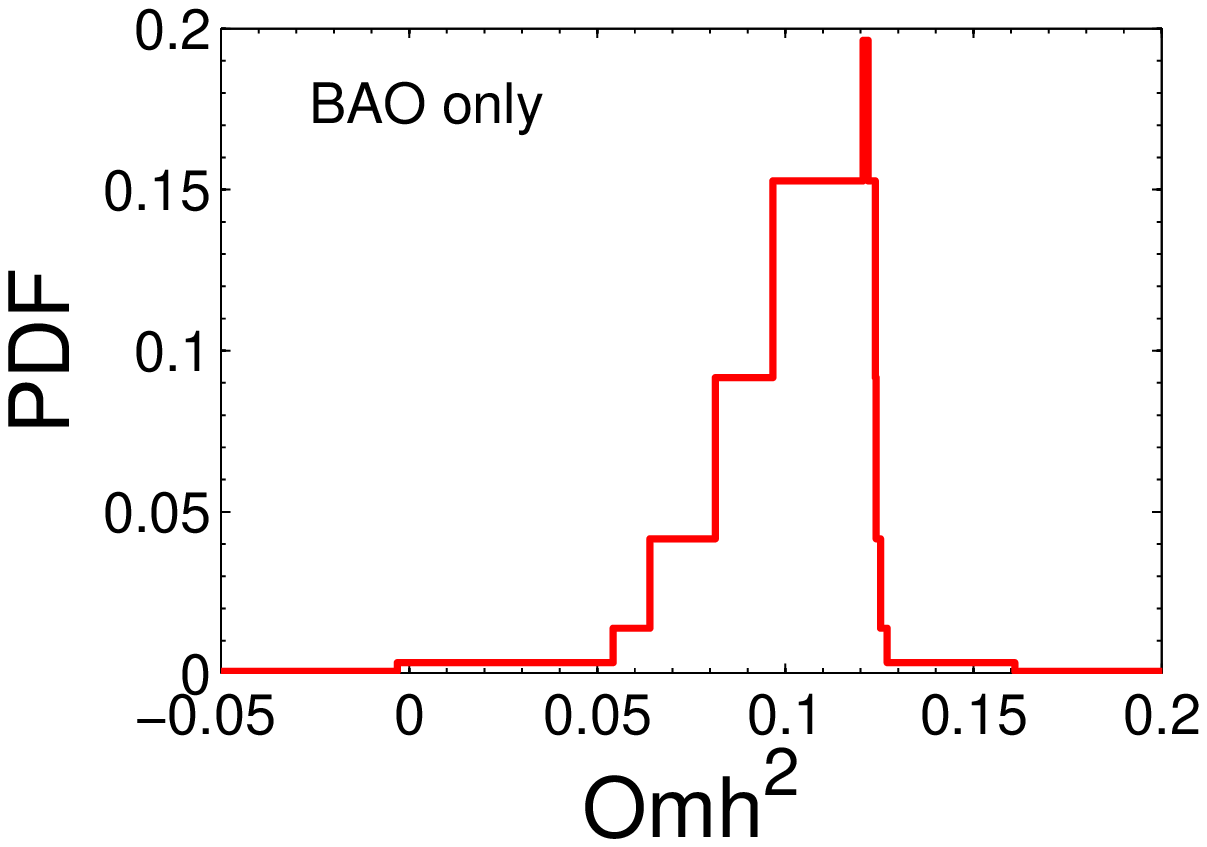}
\includegraphics[angle=0,width=40mm]{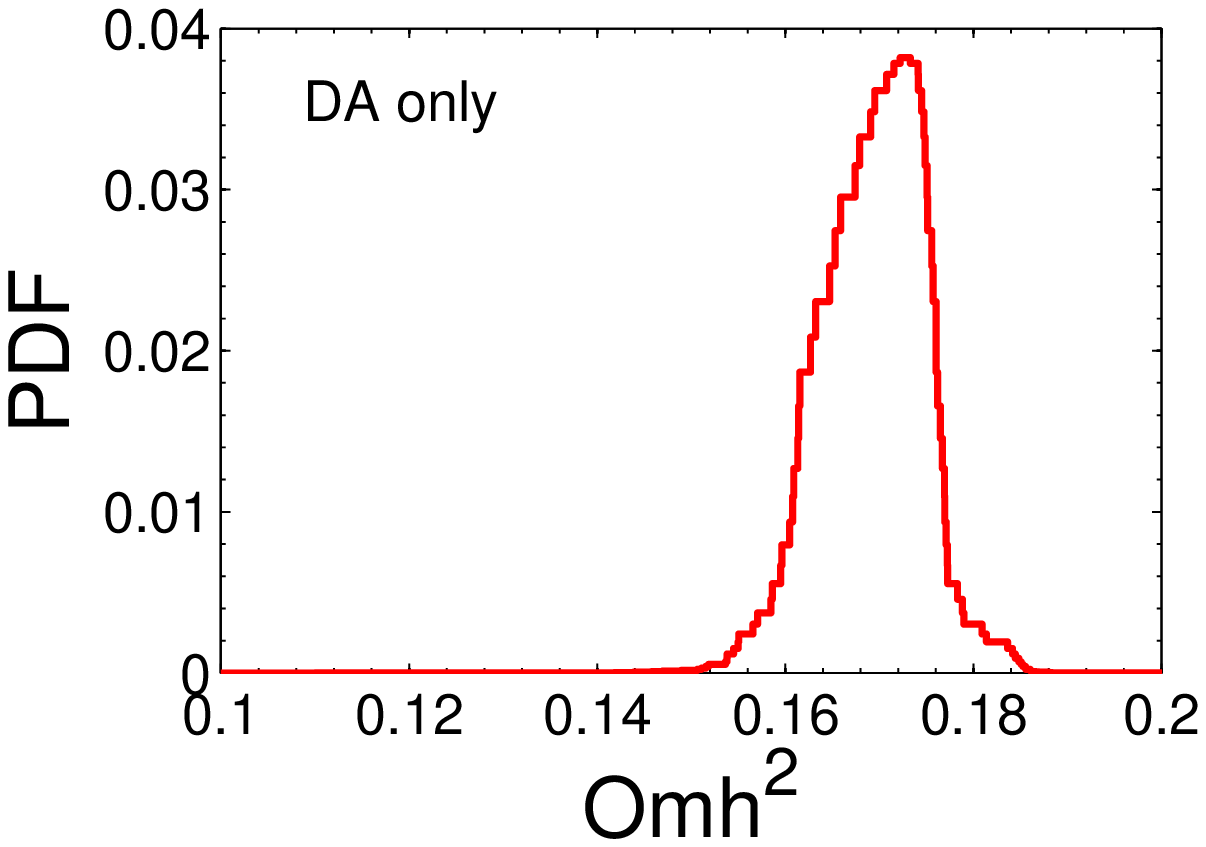}
\includegraphics[angle=0,width=40mm]{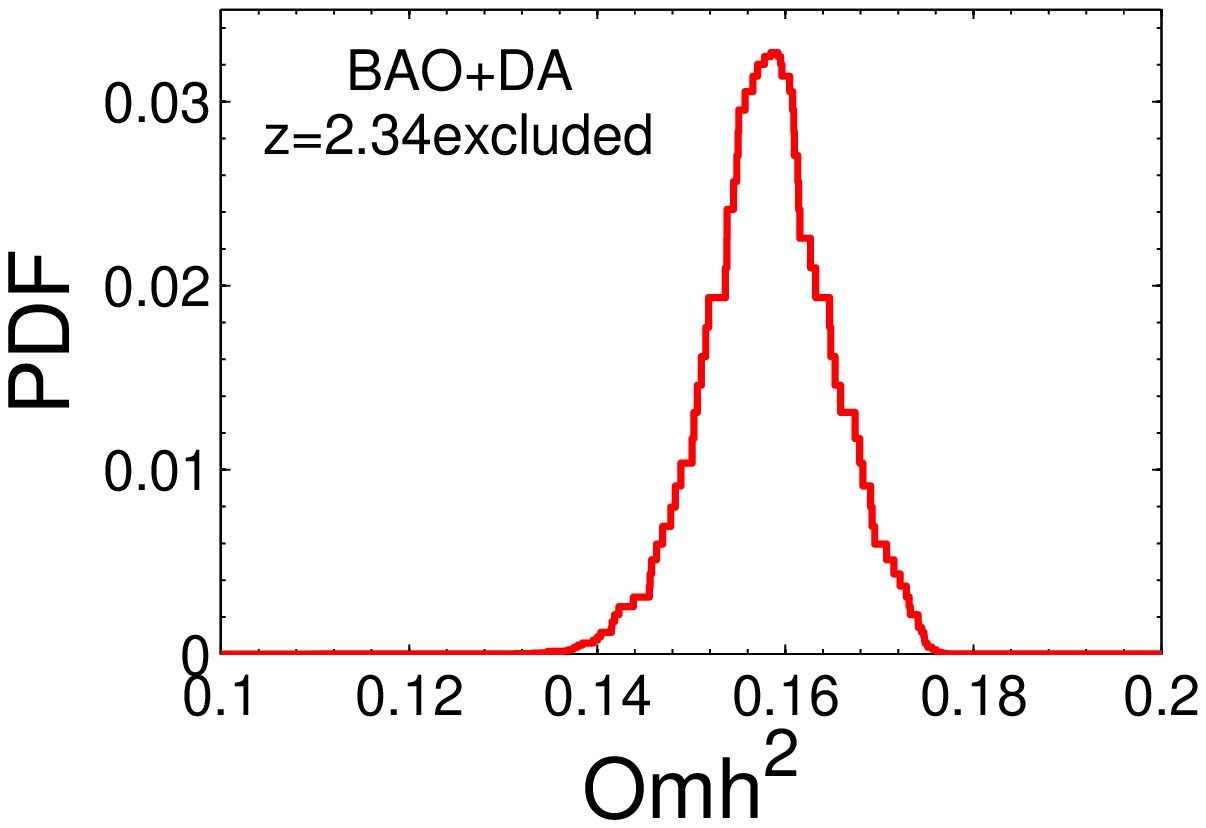}
\includegraphics[angle=0,width=40mm]{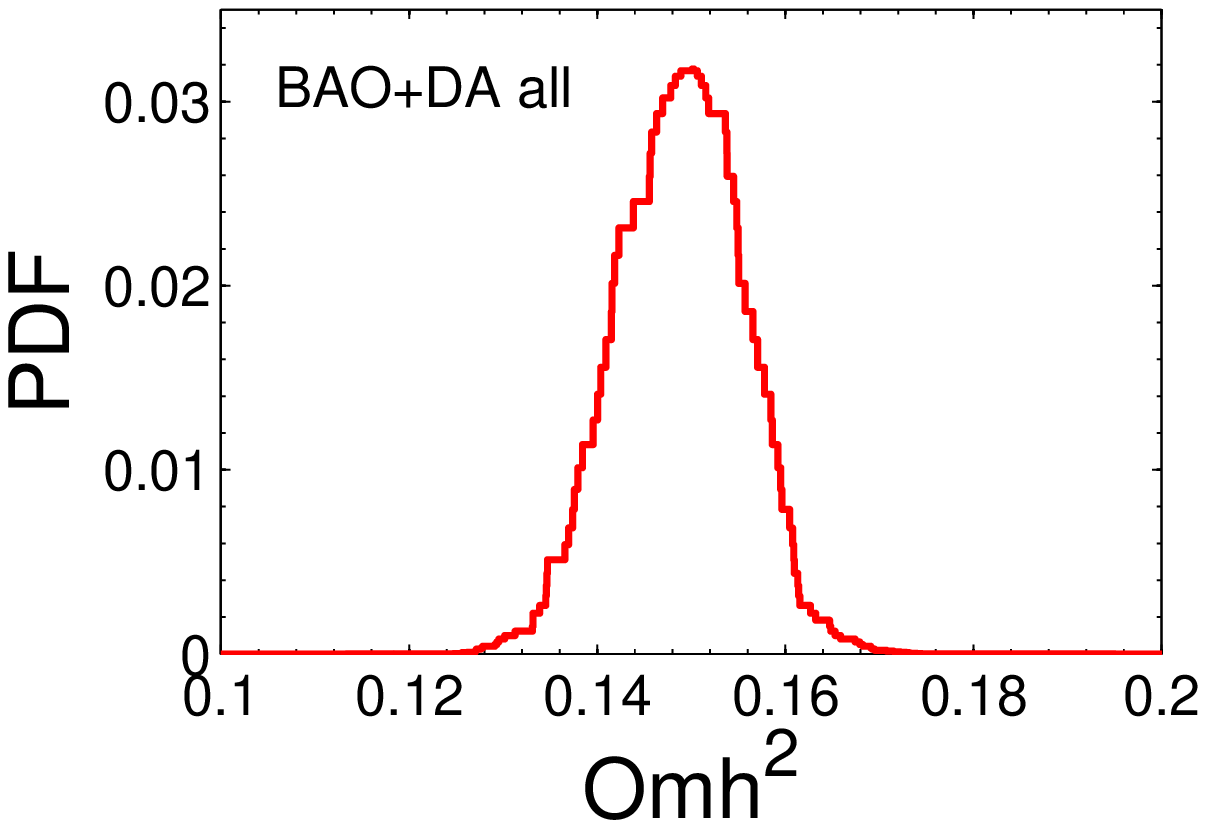}
\caption{Histograms of the $Omh^2(z_i, z_j)$ two point diagnostic calculated with different samples: $N=6$ BAO data, $N=30$ DA data, $N=35$ combined BAO+DA sample with the $H(z=2.34)$
data point excluded and the full $N=36$ combined BAO+DA data.}
\label{figPDFobomh2}
\end{center}
\end{figure*}

The results for the $Om(z_i, z_j)$ and $Omh^2(z_i, z_j)$ diagnostics obtained from the full sample and its different sub-samples are listed in Table~\ref{tableomh2} and shown on Fig.~\ref{figom} and Fig.~\ref{figomh2}.
The weighted mean approach is meaningful only under assumption of statistical independence of the data, lack of systematic effects and Gaussian distribution of errors. Hence, in order to test the Gaussianity of error distributions, we follow the approach of \citet{Chen}, \citet{RatraMedian} and \citet{Crandall}. Their idea was to construct an error distribution, a histogram of measurements as a function of $N_{\sigma}$, the number of standard deviations that a measurement deviates from a central estimate. For example, the $N_{\sigma}$ for the $Om(z_i, z_j)$ observable with respect to its weighted mean value would be: 
$N_{\sigma,k} = (Om(z_i, z_j) - Om_{(w.m.)})/ \sigma_{Om,ij}$ where $k$ -- index identifies the pair $(i,j)$. In a similar manner we calculate $N_{\sigma}$ with respect to the median value $N_{\sigma,k} = (Om(z_i, z_j) - Om_{(m.s.)})/ \sigma_{Om,ij}$. The percentage of measurements having $|N_{\sigma}|<1$ is a convenient measure of deviation from the Gaussian distribution, for which it  should be equal to $68.3\;\%$.
Therefore, in Table~\ref{tableomh2} (and also later in Tables~\ref{tableomR} and ~\ref{tableomh2R}) we report corresponding percentage of the distribution falling within $\pm 1\sigma$ i.e. $|N_{\sigma}|<1$. One clearly sees that they strongly deviate from the Gaussian expectation. We also performed the Kolmogorov-Smirnov test which strongly rejected the hypothesis of Gaussianity in each sub-sample (with p-values ranging from $10^{-4}$ to $10^{-7}$). Therefore we can conclude that the weighted average scheme is not appropriate here and the median statistics is more reliable.

Results of $Om(z_i, z_j)$ shown in Table~\ref{tableomh2} and on Figure~\ref{figom} suggest that weighted mean of this diagnostic is compatible with the $\Lambda$CDM irrespective of the sample used. However, as we argued the weighted mean is not an appropriate measure in light of non-Gaussian error distribution. On the other hand, at the level of the median statistics, it is incompatible with the $\Lambda$CDM. This conclusion seems much more justified than the previous one drawn from the weighted mean. However, one can also see that the median of $Om(z_i, z_j)$ from BAO is positive and the median from DA is negative. In other words, BAO median statistics of $Om(z_i, z_j)$ two-point diagnostic suggest the quintessence ($w>-1$) while DA median suggests phantom behaviour ($w<-1$). Of course combined data inherit the DA behavior because the median is robust against ``outliers''(here, the less numerous BAO sample). One should treat these diverging conclusions as an indication of a systematic difference between BAO and DA data concerning $H(z)$ measurements.

Table~\ref{tableomh2} and Figure~\ref{figomh2} show also the results on $Omh^2(z_i, z_j)$ diagnostics. Here one can clearly see incompatibility with the  $\Lambda$CDM when the $\Omega_{m,0}h^2$ value suggested by Planck is taken as a reference. DA and BAO+DA combined data with $H(z=2.34)$
data point excluded are compatible with the $\Lambda$CDM for the weighted mean, but our previous comments raising doubts about the appropriateness of this approach are valid here as well. One can also notice the difference between BAO and DA: the $Omh^2(z_i, z_j)$ inferred from BAO is lower and the one inferred from DA is higher than the reference value. So we can conclude that even though there are systematic differences between BAO and DA both datasets of $H(z)$ measurements are not consistent with the $\Lambda$CDM.

\begin{figure*}
\begin{center}
\centering
\includegraphics[angle=0,width=40mm]{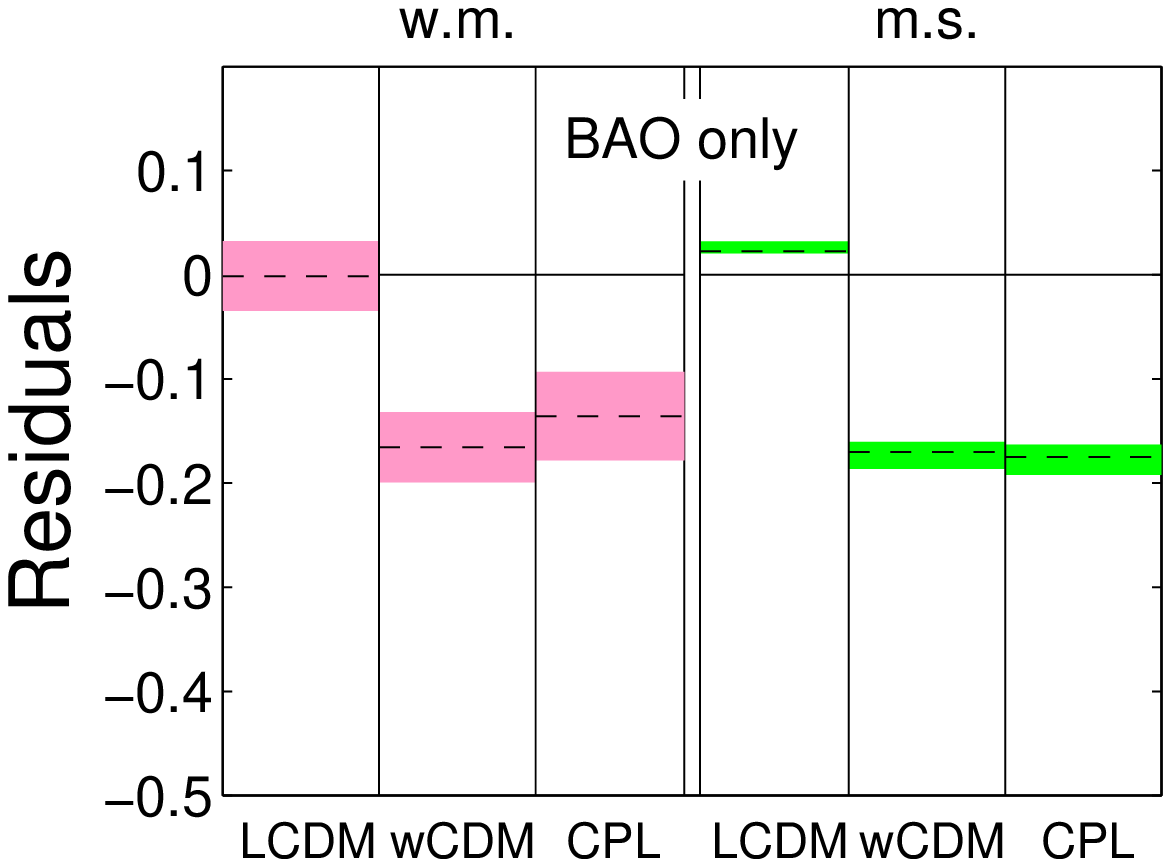}
\includegraphics[angle=0,width=40mm]{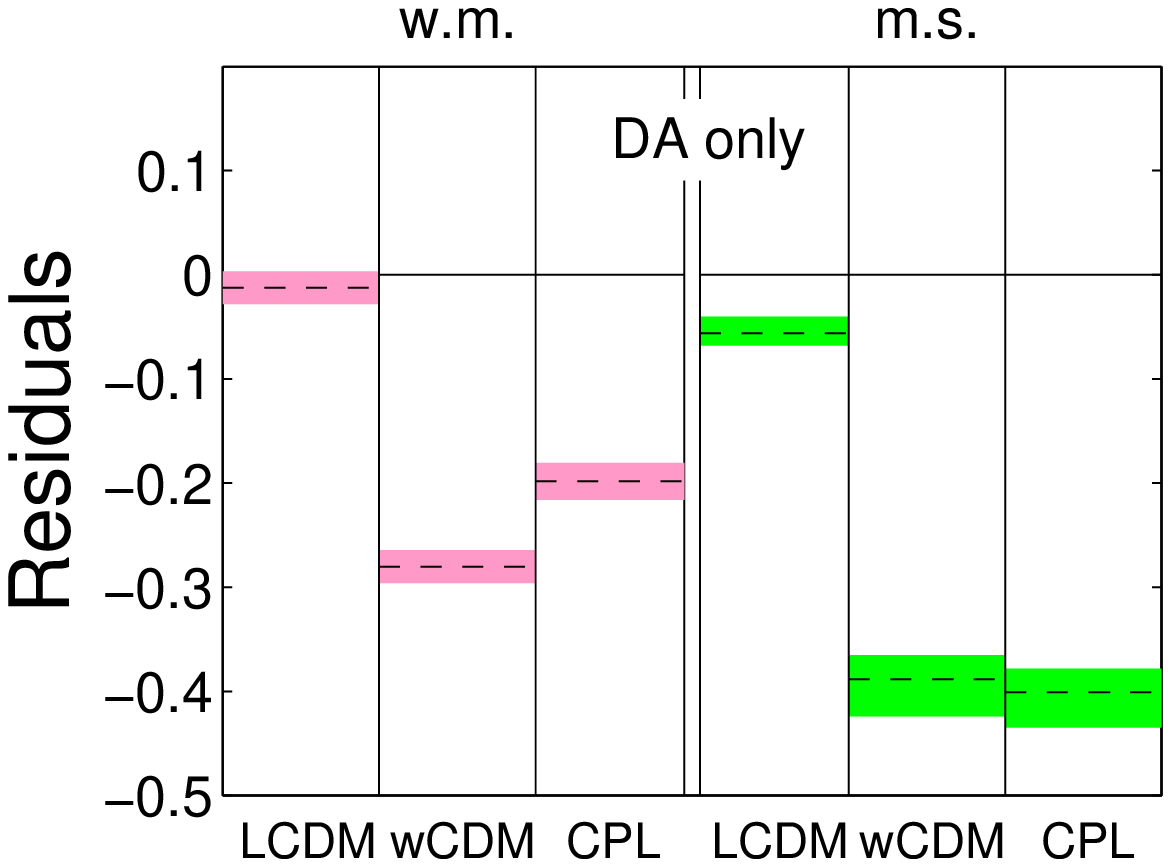}
\includegraphics[angle=0,width=40mm]{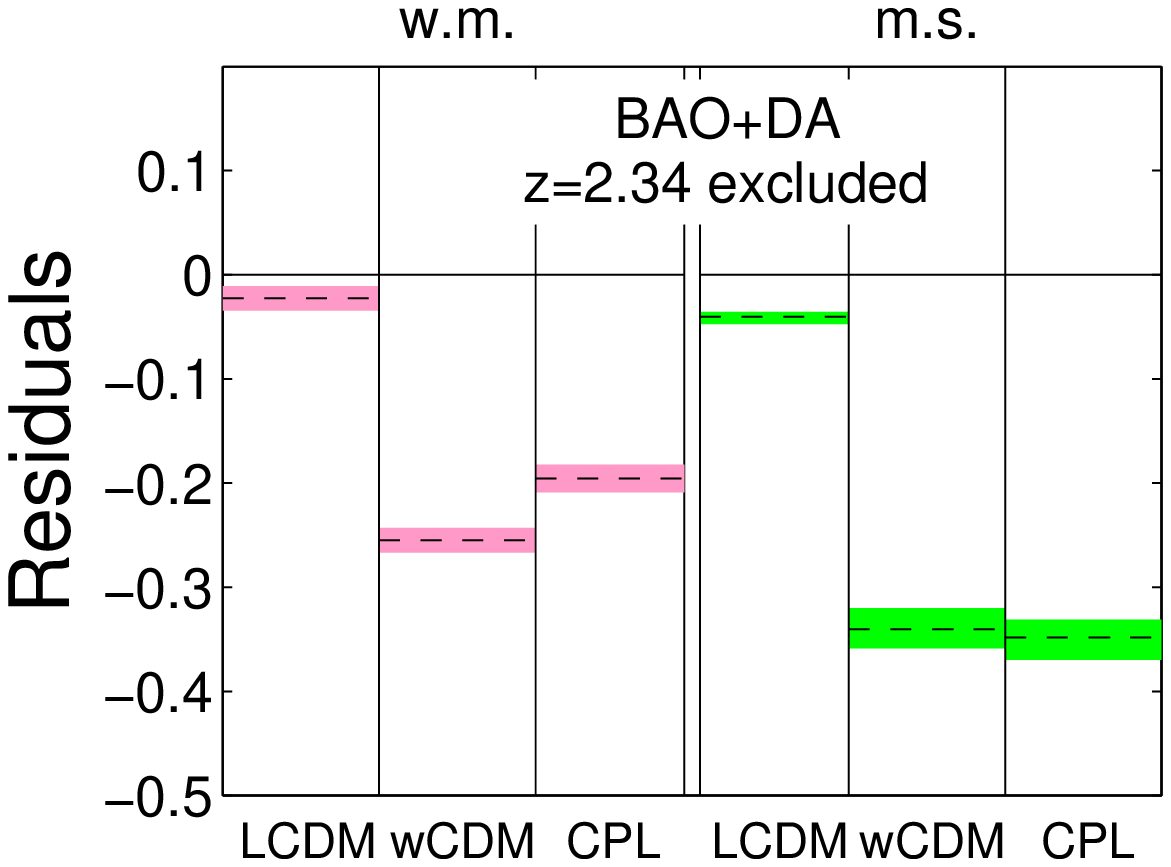}
\includegraphics[angle=0,width=40mm]{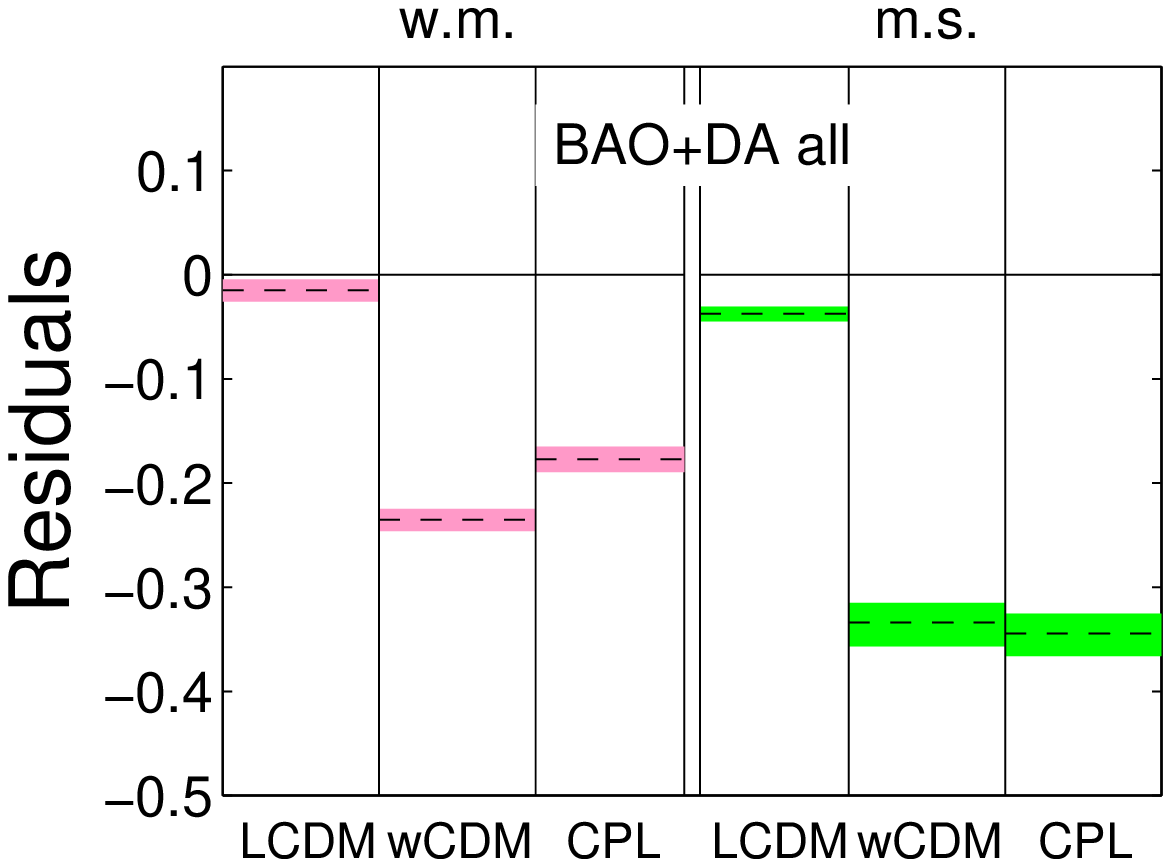}
\caption{\label{figomR}
The $Om(z_i, z_j)$ two point diagnostic residuals $R_{Om}(z_i, z_j)$ displayed as the weighted mean (left panels) and as the median value (right panels) indicated by dashed lines surrounded by color bands denoting $68\%$ confidence regions. In each panel the results for three different cosmological models are shown.
Long solid line shows the $R_{Om}(z_i, z_j)=0$ level expected for the perfect agreement between the data and the model. Four figures correspond to four respective sub-samples: $N=6$ BAO data, $N=30$ DA data, $N=35$ combined BAO+DA sample with the $H(z=2.34)$
data point excluded and the full $N=36$ combined BAO+DA data.
}
\end{center}
\end{figure*}

\begin{figure*}
\begin{center}
\centering
\includegraphics[angle=0,width=40mm]{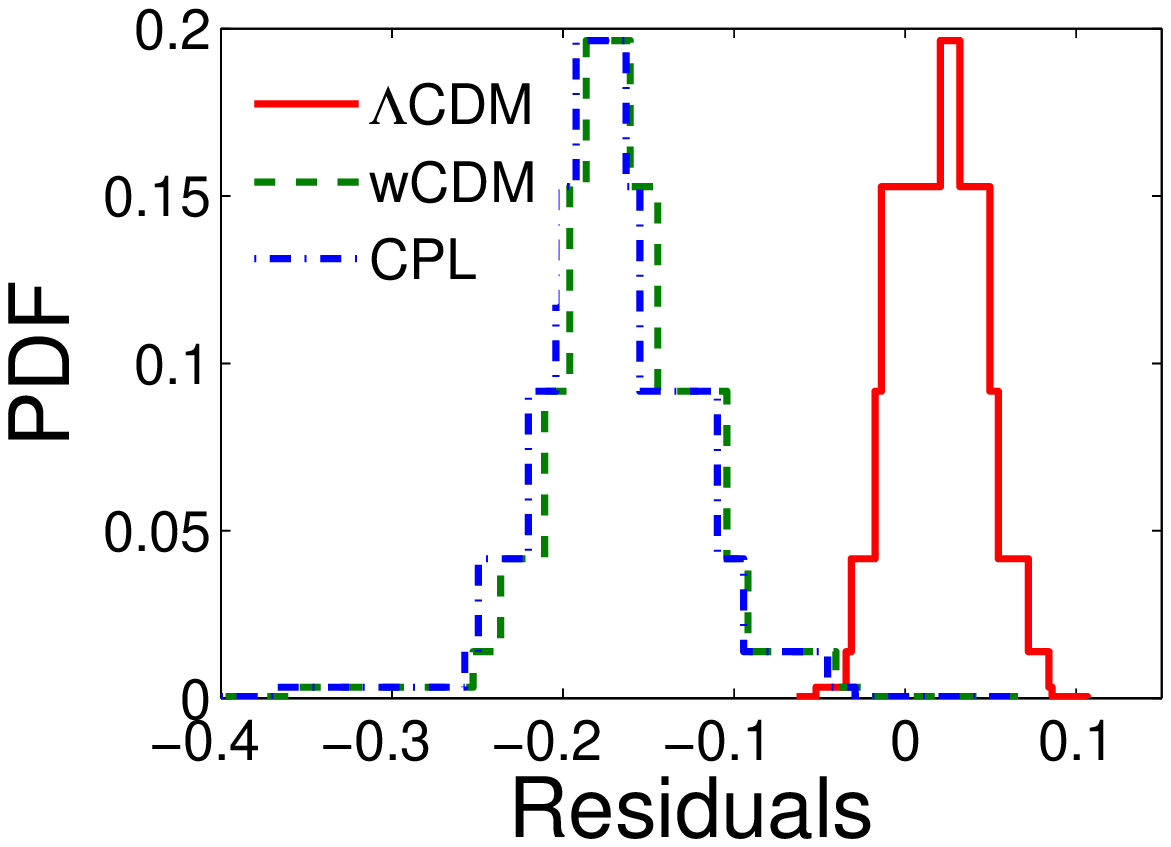}
\includegraphics[angle=0,width=40mm]{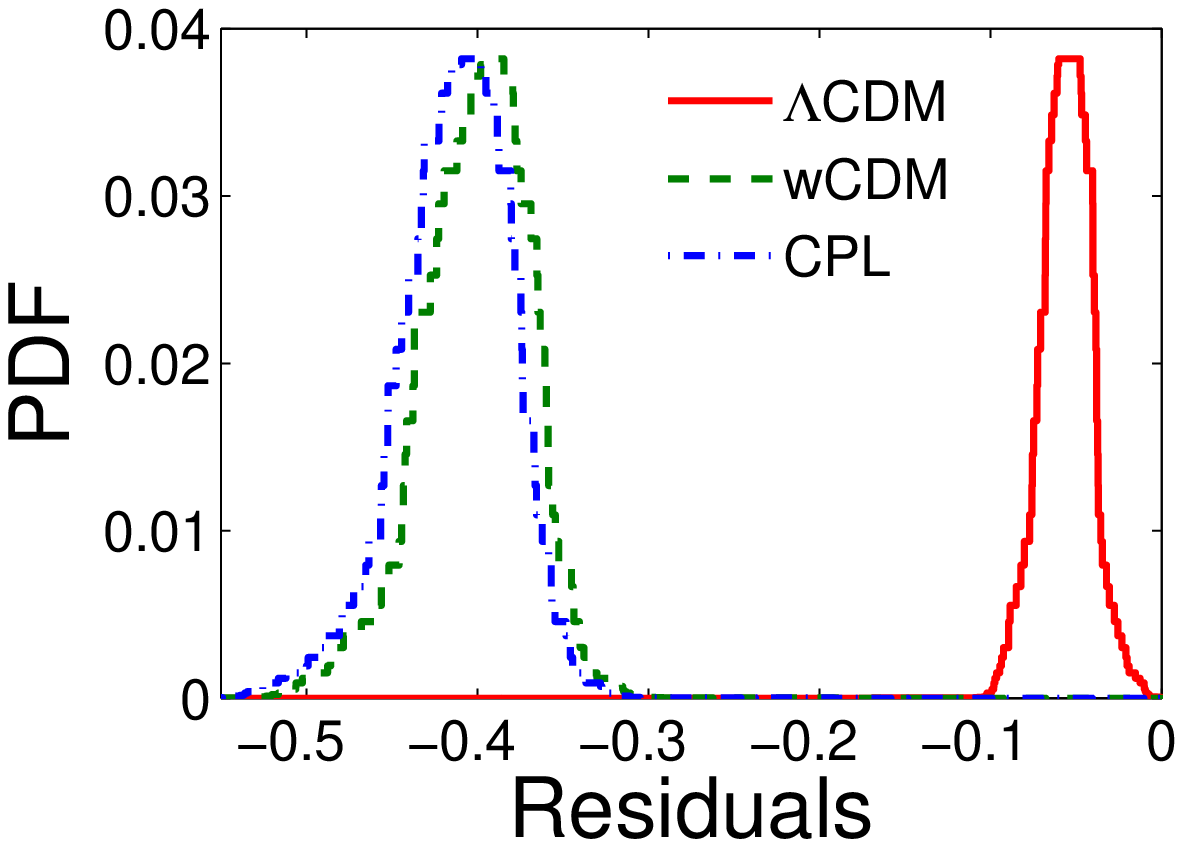}
\includegraphics[angle=0,width=40mm]{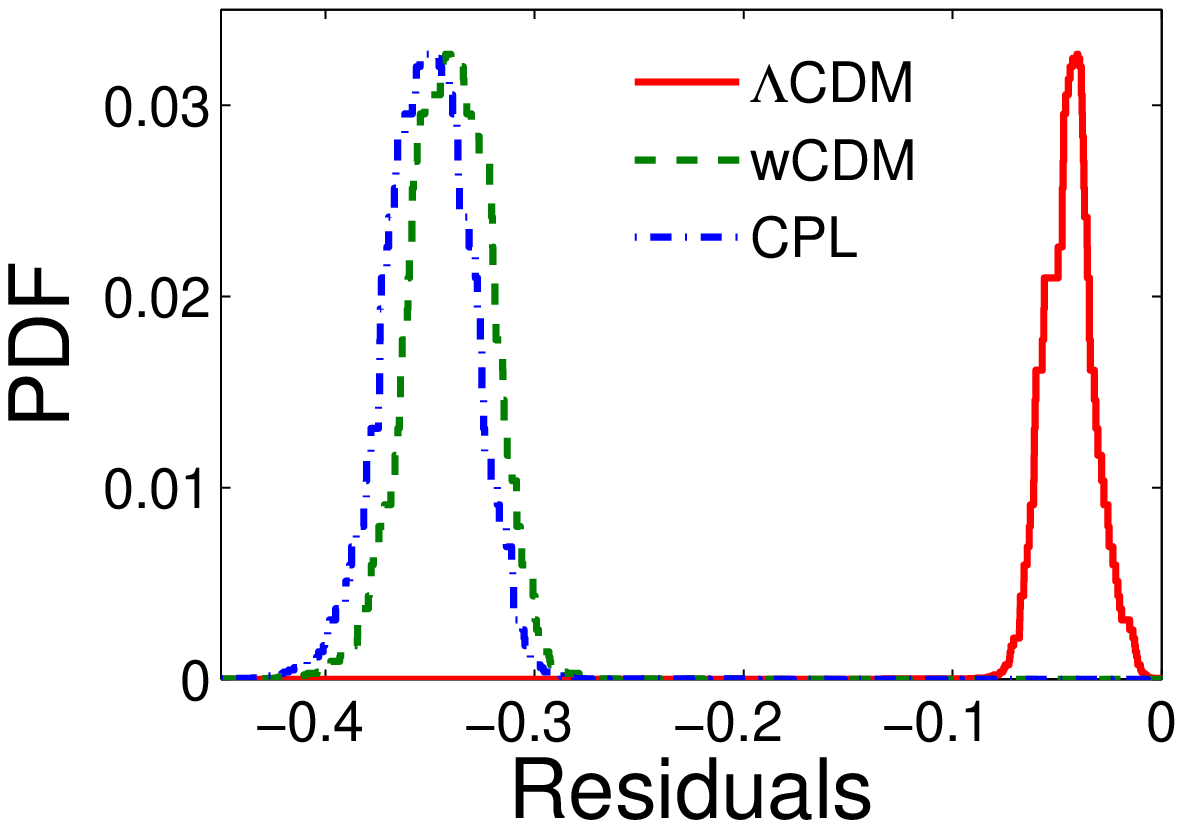}
\includegraphics[angle=0,width=40mm]{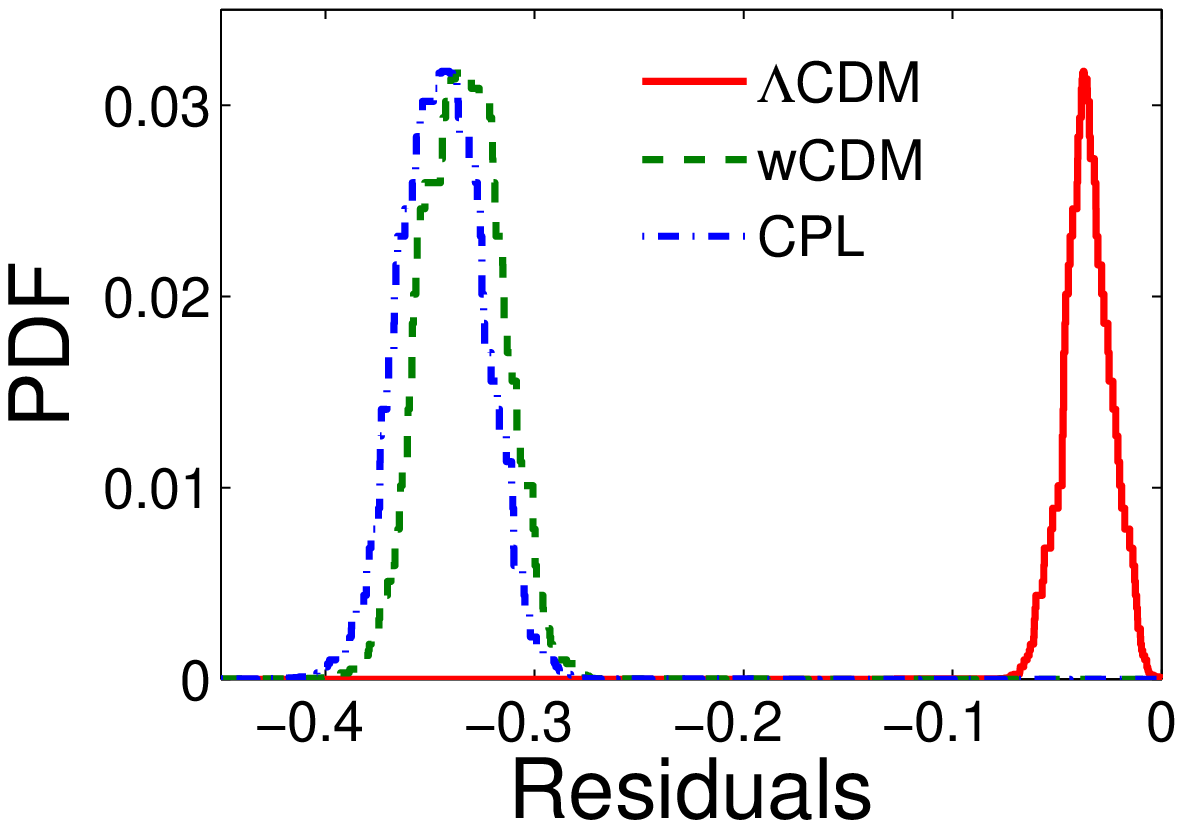}
\caption{\label{figPDFomR}
Histograms of the $Om(z_i, z_j)$ two point diagnostic residuals $R_{Om}(z_i, z_j)$ calculated with different samples: $N=6$ BAO data, $N=30$ DA data, $N=35$ combined BAO+DA sample with the $H(z=2.34)$
data point excluded and the full $N=36$ combined BAO+DA data. In each panel the results for three different cosmological models are shown.
}
\end{center}
\end{figure*}

\begin{figure*}
\begin{center}
\centering
\includegraphics[angle=0,width=40mm]{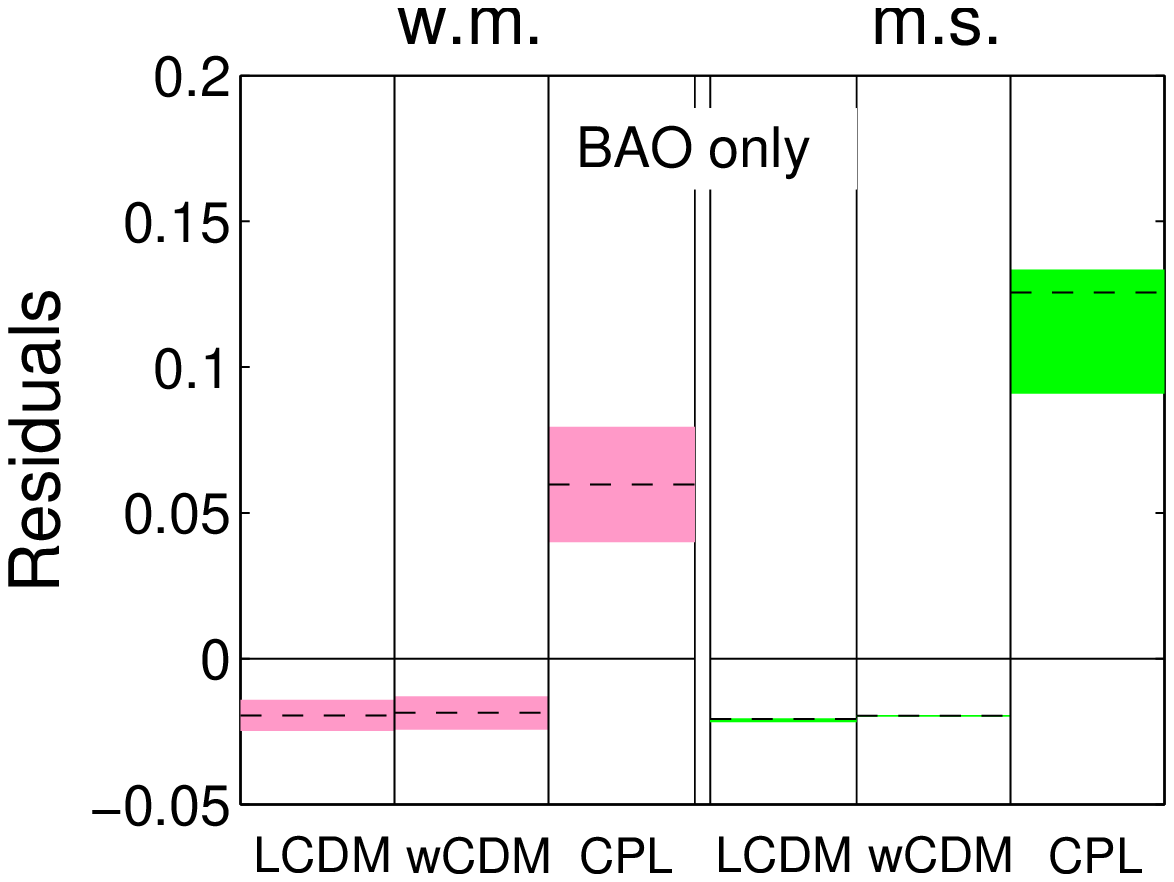}
\includegraphics[angle=0,width=40mm]{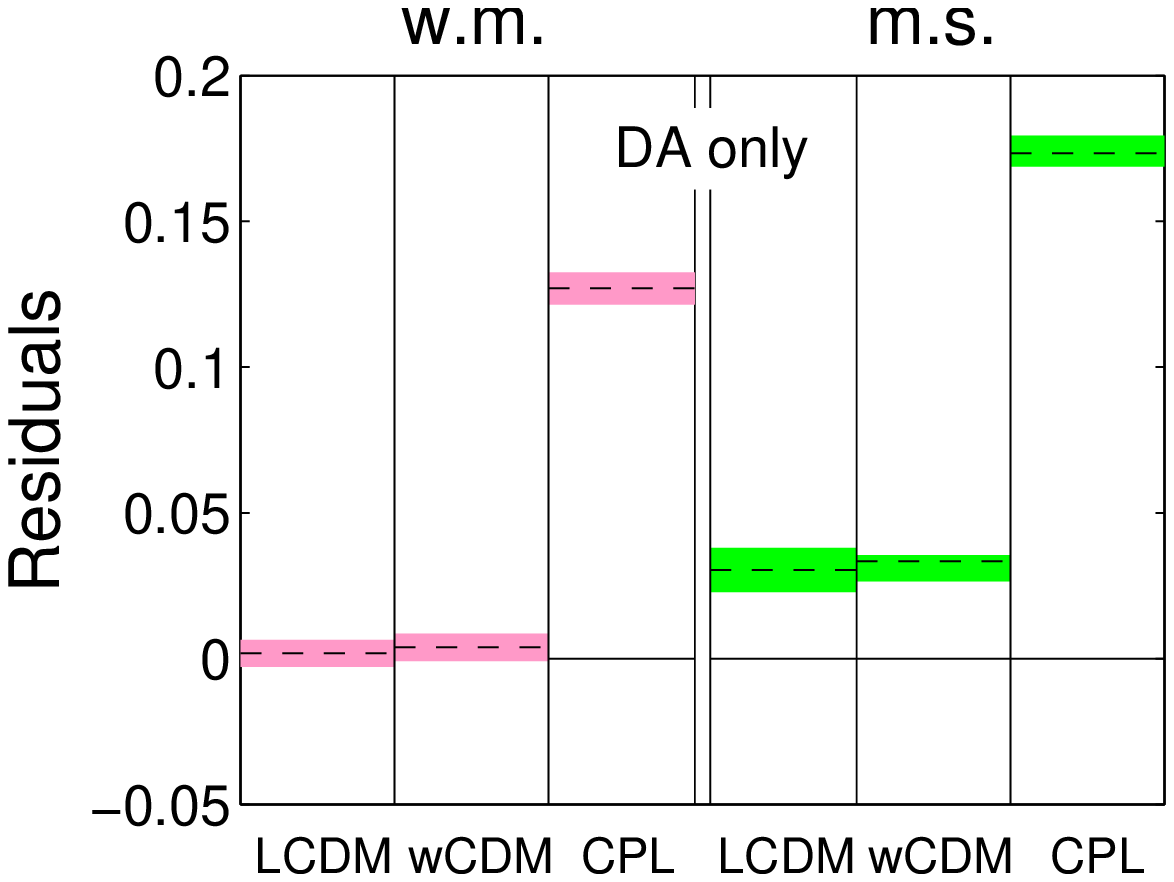}
\includegraphics[angle=0,width=40mm]{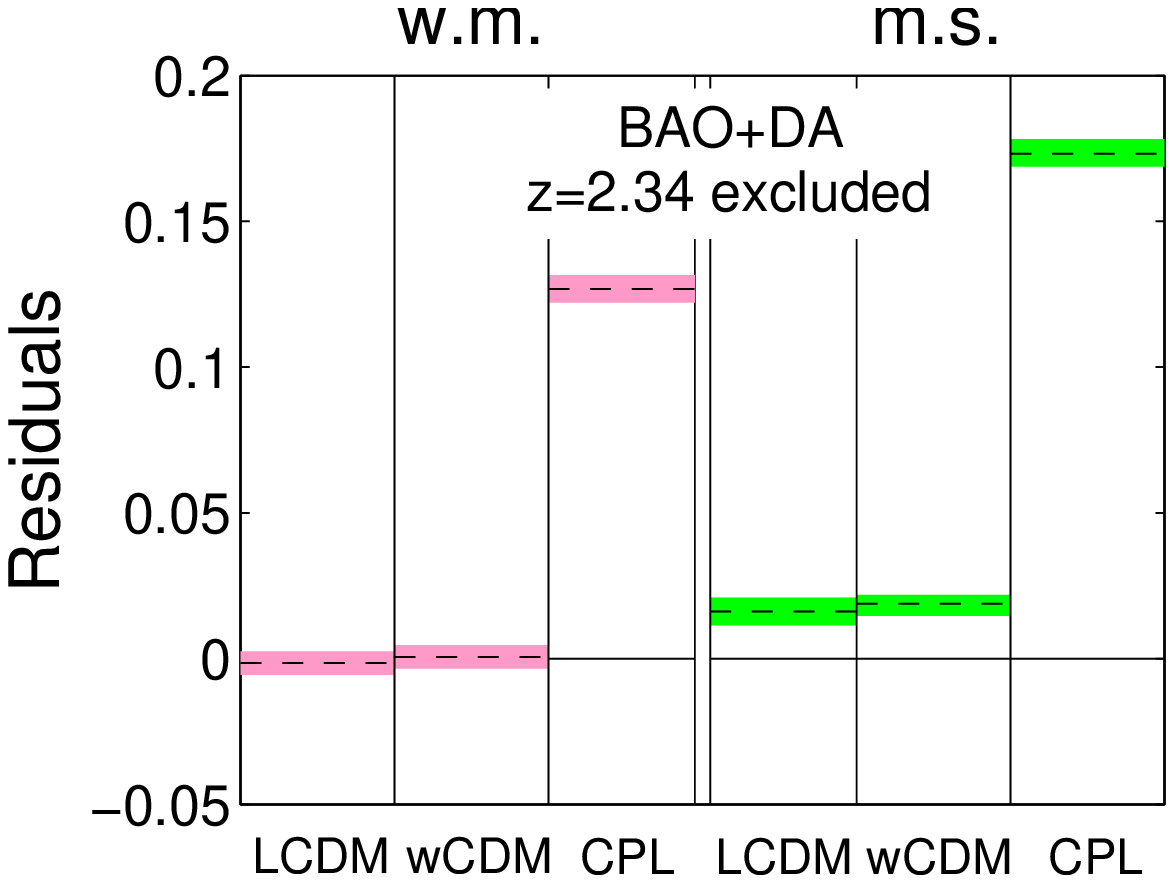}
\includegraphics[angle=0,width=40mm]{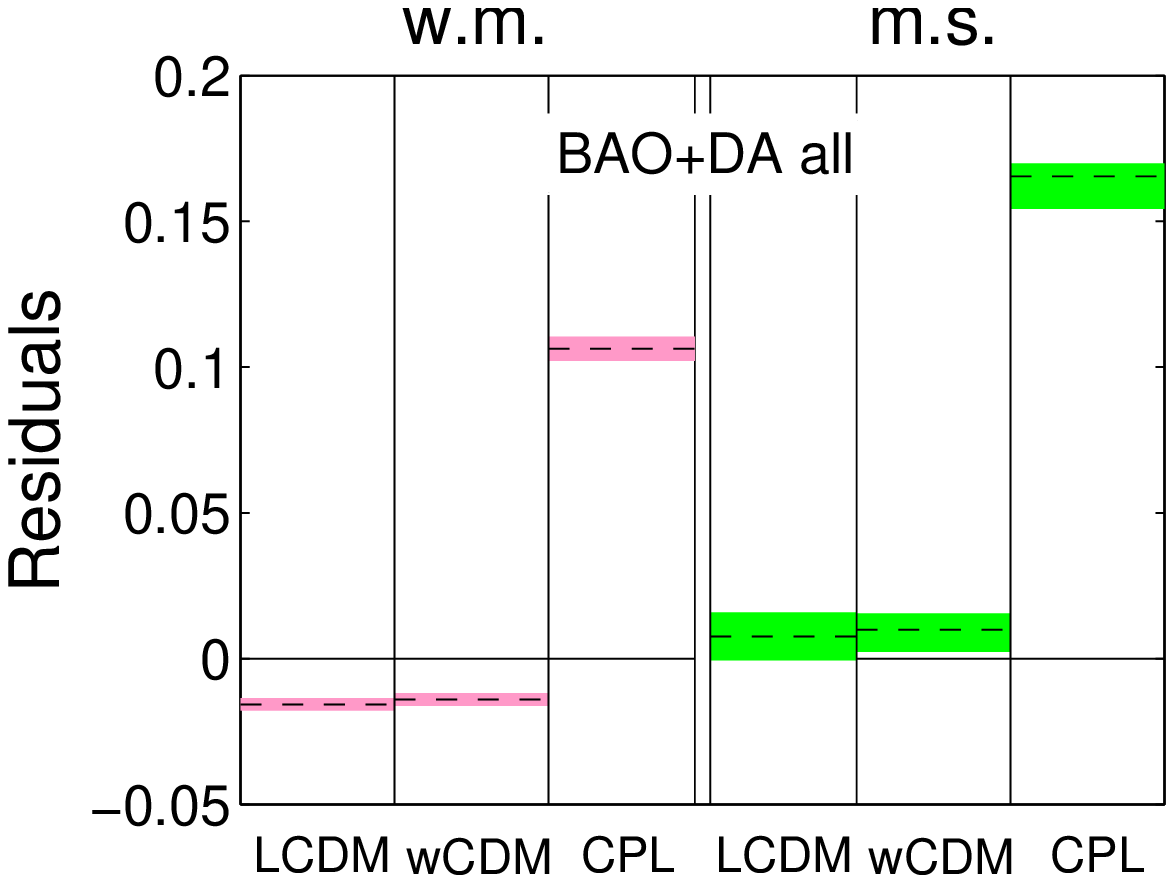}
\caption{\label{figomh2R}
The $Omh^2(z_i, z_j)$ two point diagnostic residuals $R_{Omh^2}(z_i, z_j)$ displayed as the weighted mean (left panels) and as the median value (right panels) indicated by dashed lines surrounded by color bands denoting $68\%$ confidence regions. In each panel the results for three different cosmological models are shown.
Long solid line shows the $R_{Omh^2}(z_i, z_j)=0$ level expected for the perfect agreement between the data and the model. Four figures correspond to four respective sub-samples: $N=6$ BAO data, $N=30$ DA data, $N=35$ combined BAO+DA sample with the $H(z=2.34)$
data point excluded and the full $N=36$ combined BAO+DA data.}
\end{center}
\end{figure*}

\begin{figure*}
\begin{center}
\centering
\includegraphics[angle=0,width=40mm]{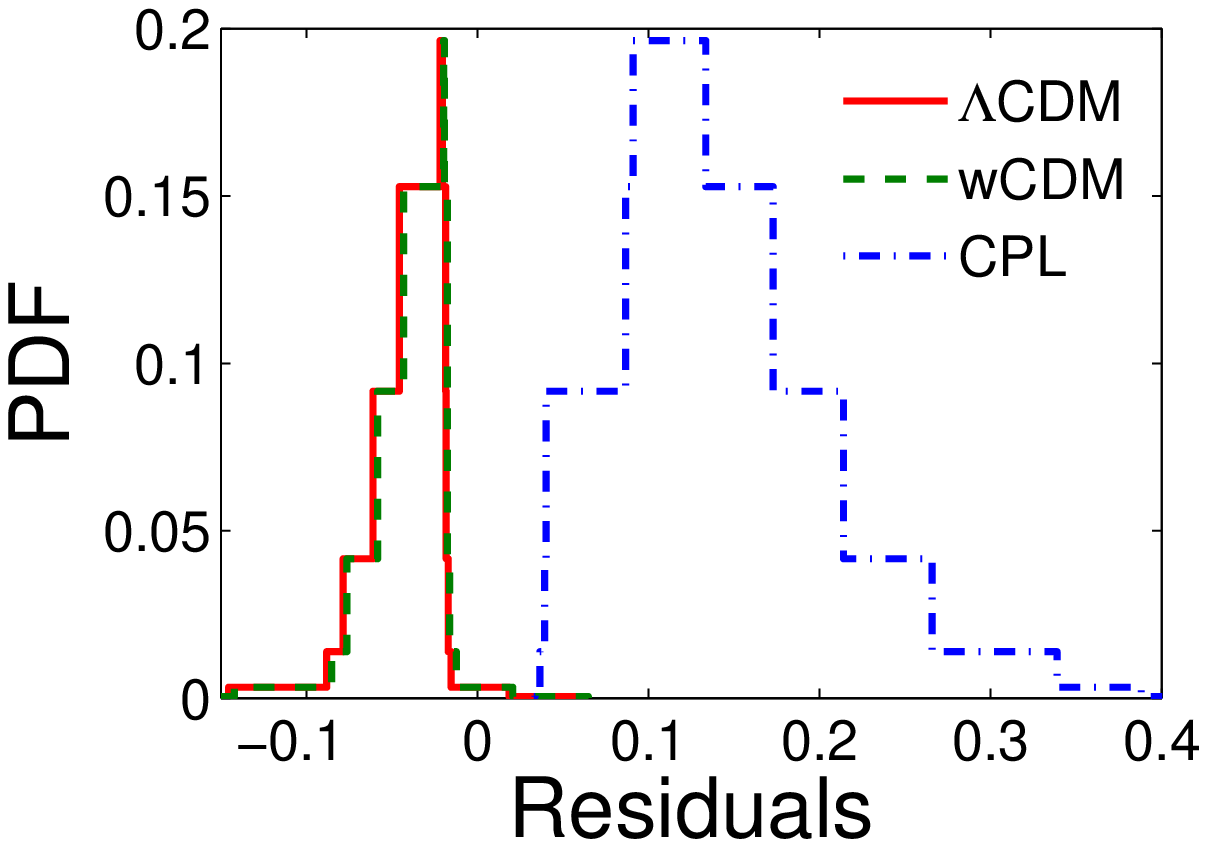}
\includegraphics[angle=0,width=40mm]{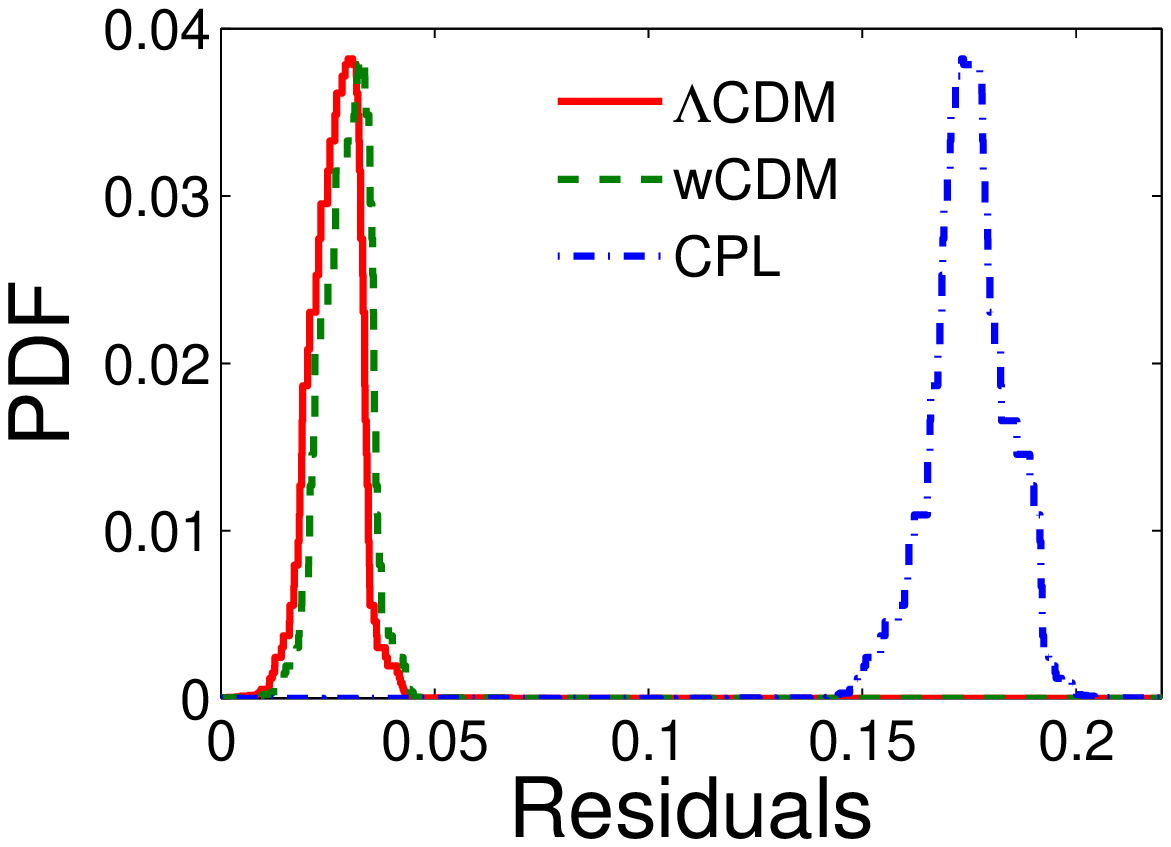}
\includegraphics[angle=0,width=40mm]{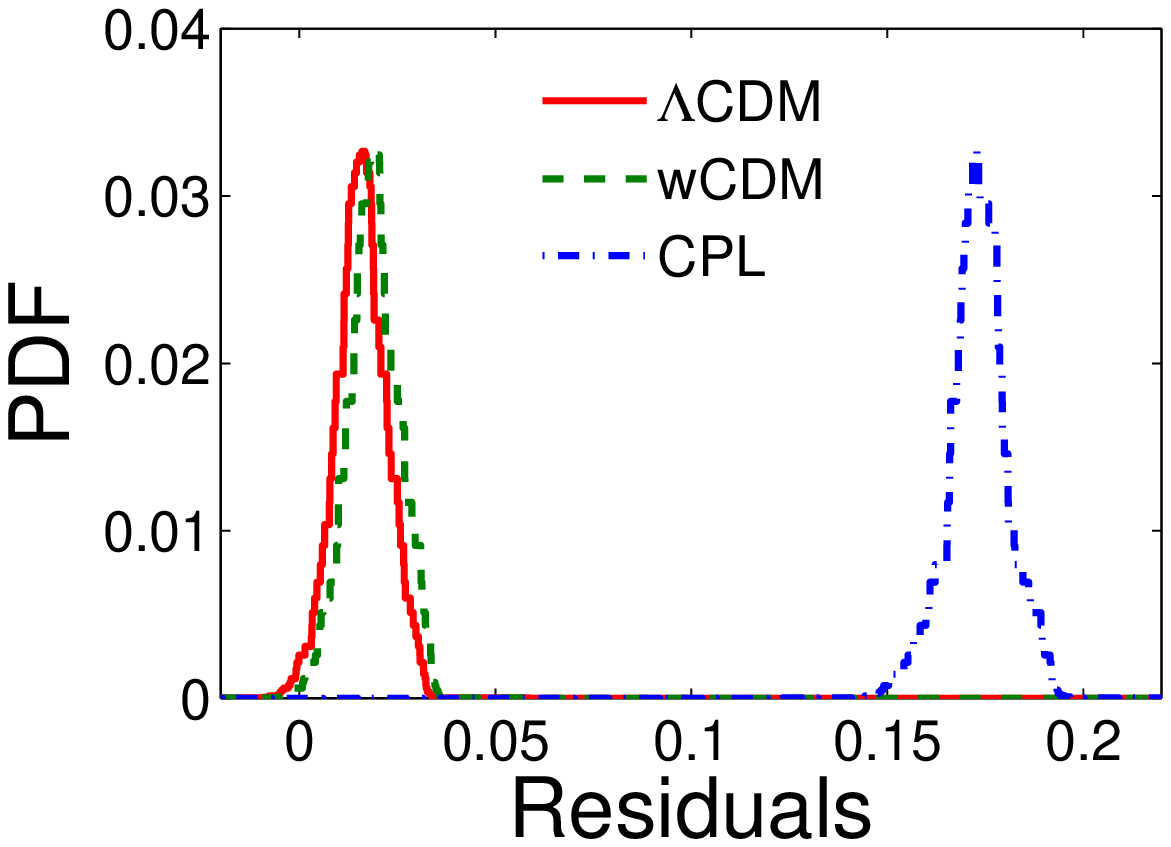}
\includegraphics[angle=0,width=40mm]{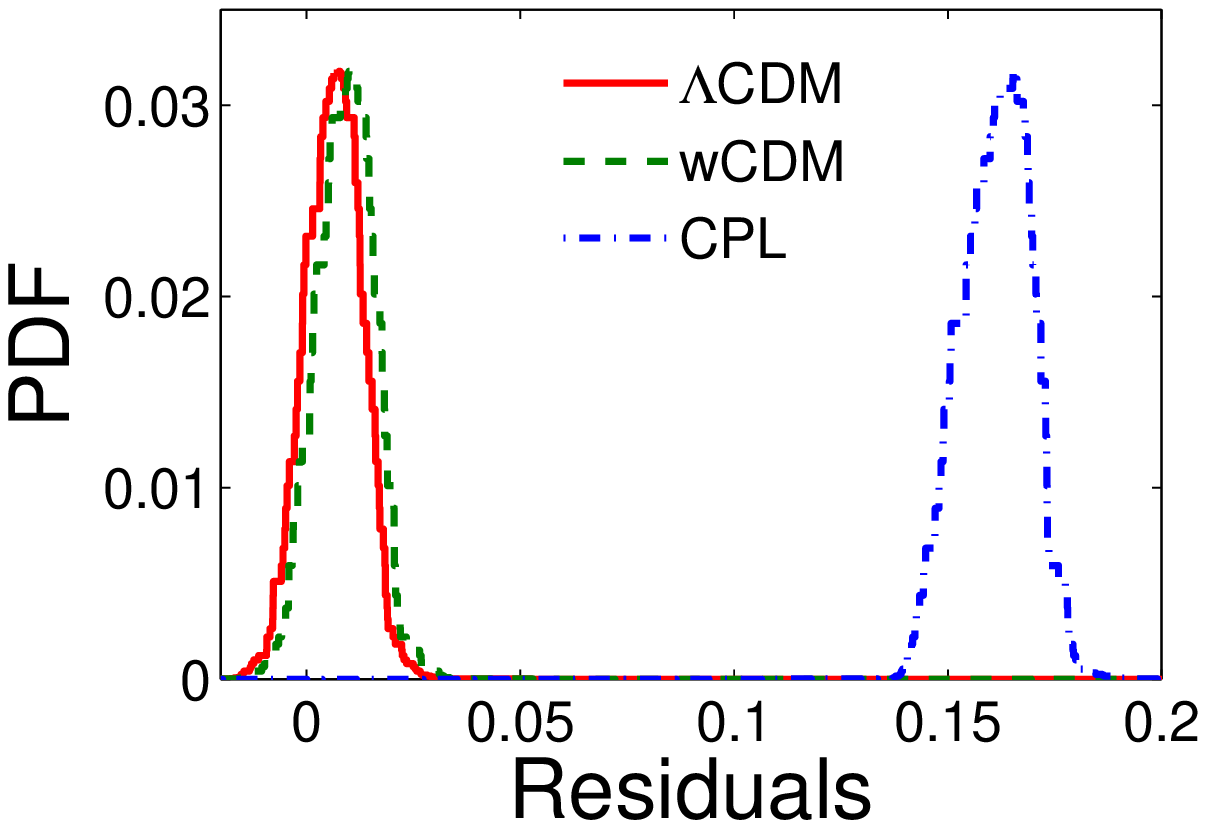}
\caption{\label{figPDFomh2R}
Histograms of the $Omh^2(z_i, z_j)$ two point diagnostic residuals $R_{Omh^2}(z_i, z_j)$ calculated with different samples: $N=6$ BAO data, $N=30$ DA data, $N=35$ combined BAO+DA sample with the $H(z=2.34)$
data point excluded and the full $N=36$ combined BAO+DA data. In each panel the results for three different cosmological models are shown.
}
\end{center}
\end{figure*}

Therefore, we can ask if some other parametrization of dark energy can perform better. In particular, we consider the simplest extensions of the $\Lambda$CDM, i.e. $wCDM$ and $CPL$ parametrization. In these models the expected values of two-point diagnostics are no longer constant, but rather the functions of redshifts given by Eqs.~(\ref{Om_wCDM}),(\ref{Om_CPL}),(\ref{Omh2_wCDM}),(\ref{Omh2_CPL}), hence we have evaluated theoretically expected values $Om(z_i, z_j)_{th}$ and $Omh^2(z_i, z_j)_{th}$ (i.e. the right hand sides of the respective equations) assuming cosmological parameters reported in Table~\ref{tableJLAp} and then we calculated the residuals $R_{Om}(z_i, z_j) = Om(z_i, z_j) - Om(z_i, z_j)_{th}$ (similarly $R_{Omh^2}(z_i, z_j)$ for the second two-point diagnostic). In principle the residuals should be zero (or rather compatible with zero in a statistical sense). If for a given model they deviate from zero more than for $\Lambda$CDM, it means that this model is less supported by $H(z)$ data in terms of two-point diagnostics.
We have summarized the residuals as the weighted mean:
\begin{equation}
R_{(w.m.)}=\frac{\sum^{n-1}_{i=1}\sum^{n}_{j=i+1}R(z_i,z_j)/\sigma^2_{R,ij}}{\sum^{n-1}_{i=1}\sum^{n}_{j=i+1}1/\sigma^2_{R,ij}}
\end{equation}\label{eqOmh2Rwm}
with the variance
\begin{equation}
\sigma^2_{R_{(w.m.)}}= \left( \sum^{n-1}_{i=1}\sum^{n}_{j=i+1}1/\sigma^2_{R,ij} \right)^{-1}
\end{equation}\label{eqOmh2Rwmsig}
and the median. The results are listed in
Table~\ref{tableomR} and Table~\ref{tableomh2R} and are shown in Figure~\ref{figomR} and Figure~\ref{figomh2R}.
Figure~\ref{figPDFomR} and Figure~\ref{figPDFomh2R} display the histograms of residuals.
Let us recall that cosmological model parameters used for calculating theoretically expected diagnostics were taken after JLA study \citep{Betoule14} as indicated in Table~\ref{tableJLAp}. Therefore, here the expected value of $Omh^2(z_i,z_j)$ in $\Lambda$CDM was not $\Omega_{m,0}h^2$ after Planck \citep{Planck2014} but the respective value suggested by the Table~\ref{tableJLAp}. Similarly, $Om(z_i,z_j)$ which is expected to vanish in the $\Lambda$CDM was calculated with $H_0$ suggested by the JLA study, not by Planck.

\begin{table*}[htp]
\caption{Results of $Om(z_i,z_j)$ two point diagnostics residuals calculated for three cosmological models: $\Lambda$CDM, wCDM and CPL on different sub-samples using the weighted mean and the median statistics.
The percentage of residuals distribution falling within $|N_{\sigma}|<1$ for the main sample and different sub-samples is shown as an indicator of non-Gaussianity.}
\begin{center}
{{\scriptsize
\begin{tabular}{l c c c c c c} \hline\hline
 $Sample/R_{(w.m.)}$ & $R_{(w.m.)(\Lambda CDM)}$ & $|N_{\sigma}|<1$ & $R_{(w.m.)(wCDM)}$ & $|N_{\sigma}|<1$ & $R_{(w.m.)(CPL)}$ & $|N_{\sigma}|<1$ \\ \hline 
Full sample (n=36) & $-0.0150\pm0.0107$ & $92.06\%$ & $-0.2354\pm0.0108$ & $68.10\%$ & $-0.1773\pm0.0122$  & $75.24\%$\\
z=2.34 excluded (n=35) & $-0.0226\pm0.0118$ & $92.27\%$ & $-0.2550\pm0.0119$ & $69.41\%$ & $-0.1956\pm0.0133$ & $76.30\%$\\
DA only (n=30) & $-0.0124\pm0.0159$ & $93.10\%$ & $-0.2803\pm0.0160$ & $68.74\%$ & $-0.1984\pm0.0179$ & $74.94\%$\\
BAO only (n=6) & $-0.0013\pm0.0335$ & $100\%$ & $-0.1657\pm0.0339$ & $66.67\%$ & $-0.1358\pm0.0426$ & $93.33\%$\\
\hline\hline
$Sample/R_{(m.s.)}$  & $R_{(m.s.)(\Lambda CDM)}$ & $|N_{\sigma}|<1$ & $R_{(m.s.)(wCDM)}$ & $|N_{\sigma}|<1$ & $R_{(m.s.)(CPL)}$ & $|N_{\sigma}|<1$ \\ \hline 
Full sample (n=36) & $-0.0375^{+0.0070}_{-0.0072}$ & $91.9\%$ & $-0.3341^{+0.0189}_{-0.0227}$ & $67.14\%$ & $-0.3444^{+0.0192}_{-0.0219}$ & $75.56\%$\\
z=2.34 excluded (n=35) & $-0.0403^{+0.0047}_{-0.0073}$ & $92.10\%$ & $-0.3405^{+0.0204}_{-0.0183}$ & $68.74\%$ & $-0.3483^{+0.0170}_{-0.0215}$ & $76.81\%$\\
DA only (n=30) & $-0.0561^{+0.0159}_{-0.0119}$ & $92.87\%$ & $-0.3883^{+0.0230}_{-0.0357}$ & $66.44\%$ & $-0.4011^{+0.0228}_{-0.0338}$ & $75.17\%$\\
BAO only (n=6) & $0.0226^{+0.0018}_{-0.0095}$ & $100\%$ & $-0.1704^{+0.0098}_{-0.0159}$ & $66.67\%$ & $-0.1749^{+0.0119}_{-0.0174}$ & $86.67\%$\\ 
\hline\hline
\end{tabular}}\label{tableomR}}
\end{center}
\end{table*}

\begin{table*}[htp]
\caption{Results of $Omh^2(z_i,z_j)$ two point diagnostics residuals calculated for three cosmological models: $\Lambda$CDM, wCDM and CPL on different sub-samples using the weighted mean and the median statistics.
The percentage of residuals distribution falling within $|N_{\sigma}|<1$ for the main sample and different sub-samples is shown as an indicator of non-Gaussianity.}
\begin{center}
{{\scriptsize
\begin{tabular}{l c c c c c c} \hline\hline
sample/$R_{(w.m.)}$  & $R_{(w.m.)(\Lambda CDM)}$ & $|N_{\sigma}|<1$ & $R_{(w.m.)(wCDM)}$ & $|N_{\sigma}|<1$ & $R_{(w.m.)(CPL)}$ & $|N_{\sigma}|<1$ \\ \hline 
Full sample (n=36) & $-0.0157\pm0.0021$ & $83.65\%$ & $-0.0140\pm0.0022$ & $83.81\%$ & $0.1063\pm0.0041$  & $68.73\%$ \\
z=2.34 excluded (n=35) & $-0.0016\pm0.0040$ & $82.52\%$ & $0.0006\pm0.0040$ & $82.52\%$ & $0.1268\pm0.0047$ & $76.64\%$ \\
DA only (n=30) & $0.0018\pm0.0046$ & $81.84\%$ & $0.0039\pm0.0047$ & $82.30\%$ & $0.1270\pm0.0055$ & $75.40\%$ \\
BAO only (n=6) & $-0.0194\pm0.0053$ & $100\%$ & $-0.0186\pm0.0057$ & $100\%$ & $0.0597\pm0.0197$ & $80\%$ \\ 
\hline\hline
sample/$R_{(m.s.)}$  & $R_{(m.s.)(\Lambda CDM)}$ & $|N_{\sigma}|<1$ & $R_{(m.s.)(wCDM)}$ & $|N_{\sigma}|<1$ & $R_{(m.s.)(CPL)}$ & $|N_{\sigma}|<1$ \\ \hline 
Full sample (n=36) & $0.0076^{+0.0049}_{-0.0082}$ & $79.37\%$ & $0.0099^{+0.0056}_{-0.0075}$ & $80\%$ & $0.1654^{+0.0045}_{-0.0112}$ & $70.79\%$ \\
z=2.34 excluded (n=35) & $0.0162^{+0.0029}_{-0.0048}$ & $85.04\%$& $0.0189^{+0.0031}_{-0.0042}$ & $85.55\%$ & $0.1731^{+0.0051}_{-0.0043}$ & $75.29\%$ \\
DA only (n=30) & $0.0304^{+0.0027}_{-0.0076}$ & $87.82\%$ & $0.0335^{+0.0020}_{-0.0070}$ & $88.51\%$ & $0.1733^{+0.0061}_{-0.0045}$ & $74.48\%$ \\
BAO only (n=6) & $-0.0207^{+0.0002}_{-0.0011}$ & $100\%$ & $-0.0196^{+0.0002}_{-0.0003}$ & $100\%$ & $0.1256^{+0.0079}_{-0.0347}$ & $60\%$ \\ 
\hline\hline
\end{tabular}}\label{tableomh2R}}
\end{center}
\end{table*}

One can see from the Figure~\ref{figomR} that residuals $R_{Om}(z_i, z_j)$ for the $\Lambda$CDM are closer to zero than for $wCDM$ or CPL models irrespectively of the sample. From the Table~\ref{tableomR} one can see that residuals of $wCDM$ or CPL models summarized in the weighted mean scheme are more than $15\sigma$ away from the expected value of zero (for the full sample). This deviation in terms of median statistics is even bigger. One can also see it clearly in Figure~\ref{figPDFomR} where the histograms of $R_{Om}(z_i, z_j)$ are shown. In the case of $Omh^2(z_i,z_j)$ diagnostics the performance of $\Lambda$CDM and $wCDM$ is similar: using the full sample, weighted mean of $R_{Omh^2}(z_i, z_j)$ residuals is at $7\sigma$ away from zero. As shown in Figure~\ref{figPDFomh2R} the bulk of the $R_{Omh^2}(z_i, z_j)$ distributions for $\Lambda$CDM or $wCDM$ contains zero in their tails, while the distribution for CPL model is considerably away from the zero (it corresponds to $26\sigma$ for the weighted mean).  It seems that despite there is some tension between the $\Lambda$CDM model and the two-point diagnostics evaluated on the most recent $H(z)$ data, as noticed e.g. in \citep{Sahni2014,Ding2015}, yet this model performs better than its immediate extensions: $wCDM$ or $CPL$, in particular the last one. It should be stressed that the above mentioned performance of different models refers only to the two-point diagnostics considered. Therefore, this cannot be treated as a decisive ranking of competing models. Major obstacle for using two-point diagnostics for cosmological models other than $\Lambda$CDM is that in such cases it ceases to be so strong ``screening test'' because its expected value is no longer a number, but a function of redshift involving cosmological model parameters, which should be somehow assessed prior to use of this test. 

\section{Conclusions}

Two point diagnostics: $Om(z_i,z_j)$ and $Omh^2(z_i,z_j)$ have been introduced as an interesting tool for testing the validity of the $\Lambda$CDM model. Reliable data concerning expansion rates of the Universe at different redshifts $H(z)$ are crucial for their successful application. Now we are at the moment in time when fairly reliable data of this kind are being obtained from DA and BAO techniques. Therefore, in this paper we examined both diagnostics on the  comprehensive set comprising data compiled in \citet{Ding2015} \footnote{They are essentially the same as the data from \citet{Farooq} enriched by BAO measurement of \citet{Delubac}.} supplemented by the most recent DA measurements by \citet{Moresco15} and \citet{Moresco16}.
An important motivation for this study was the paper by \citet{Sahni2014} where, based on three $H(z)$ measurements from BAO (including the $z=2.34$ measurement by \citet{Delubac}) they claimed that recent precise measurements of expansion rates
at different redshifts suggest a severe tension with the $\Lambda$CDM model. Our study \citep{Ding2015} confirmed this claim, however this was based only on one particular two-point diagnostic $Omh^2(z_i,z_j)$ which is expected to be equal to $\Omega_{m,0}h^2$ in the $\Lambda$CDM model. In this paper we not only used a bigger data set -- enriched by the most recent DA data -- but we also considered the $Om(z_i,z_j)$ two point diagnostic which is expected to be zero in the $\Lambda$CDM. Therefore this diagnostic does not depend on our knowledge of the matter density parameter and the uncertainty about its value does not propagate into the inference. Being aware that BAO and DA techniques are prone to different systematic uncertainties, and because of the big leverage of the $z=2.34$ data point we have analyzed not only full combined sample of $N=36$ BAO+DA data, but also different sub-samples. It turned out that both two-point diagnostics have non-Gaussian distributions and therefore the median statistic is more appropriate way to describe them than the weighted mean scheme. The median statistic results support the claim that $H(z)$ data seem to be in conflict with the
$\Lambda$CDM model. However, two-point diagnostics evaluated on BAO and DA data deviate in different directions from the expectations concerning $\Lambda$CDM. This indicates that there are serious systematic effects in these two approaches. The DA method is very simple and transparent in its design. The major source of systematics is the adopted population synthesis model which quantifies the relation between $D4000$ spectral break, metallicity, star formation history and the age of the galaxy \citep{Moresco16}. On the contrary, in spite of its huge statistical power, BAO technique is much more complex.
In order to derive $H(z)$ from the large-scale clustering patterns of galaxies, one has not only determine the baryon acoustic peak in angle-averaged clustering pattern but also measure the Alcock - Paczynski effect from the two-point statistics of galaxy clustering. This requires good understanding of redshift-space distortions and sophisticated statistical methods. It suggests that BAO data on $H(z)$ should be treated with caution when used for constraining cosmological model, much more than in the case of using more direct observable -- the``dilation scale'' distance $D_V(z)$.

We have also asked a question, if other cosmological models, alternative to the $\Lambda$CDM perform better. In particular we considered wCDM and CPL models. However, the diagnostic test was not so simple: we had to confront $Omh(z_i,z_j)$ and $Omh^2(z_i,z_j)$ diagnostics calculated from $H(z)$ data against theoretically expected (redshift dependent) counterparts. We performed this calculating the ``observed - expected'' residuals. It turned out that despite the revealed mismatch between the data and the $\Lambda$CDM, this model is still in better agreement with the data than wCDM or CPL. There is one caveat in our approach, namely that in order to evaluate theoretically expected counterparts of two-point diagnostics we have taken cosmological parameters best fitted by joint JLA study \citep{Betoule14} as a reference point. It would be tempting and more consistent to perform the fit of cosmological parameters based on the two-point diagnostics. However, because of the error distribution revealed on Fig.~\ref{figpair} it would not give results competitive with other techniques. On the other hand, since the systematics underlying this peculiar behaviour of uncertainties has been partly recognized it could be used to define and use suitable subsamples better suited for cosmological inference.

\section*{Acknowledgements}
This work was supported by the Ministry of Science and Technology National Basic Science Program (Project 973) under Grants Nos. 2012CB821804 and 2014CB845806, the Strategic Priority Research Program ''The Emergence of Cosmological Structure'' of the Chinese Academy of Sciences (No. XDB09000000), the National Natural Science Foundation of China under Grants Nos. 11373014 and 11073005. S.C. also acknowledges the support from China Postdoctoral Science Foundation under Grant No. 2014M550642. M.B. obtained approval of foreign talent introducing project in China and gained special fund support of foreign knowledge introducing project. He also gratefully acknowledges hospitality of Beijing Normal University.

\end{document}